\documentclass[sn-mathphys,Numbered]{sn-jnl}

\usepackage[dvipsnames]{xcolor}
\usepackage{hyperref}
\usepackage{graphicx}%
\usepackage{multirow}%
\usepackage{amsmath,amssymb,amsfonts}%
\usepackage{amsthm}%
\usepackage{mathrsfs}%
\usepackage[title]{appendix}%
\usepackage{xcolor}%
\usepackage{textcomp}%
\usepackage{manyfoot}%
\usepackage{booktabs}%
\usepackage{algorithm}%
\usepackage{algorithmicx}%
\usepackage{algpseudocode}%
\usepackage{listings}%

\theoremstyle{thmstyleone}%
\theoremstyle{thmstyletwo}%

\theoremstyle{thmstylethree}%

\begin{document}

\title[Polynomial Solutions...]{Polynomial Solutions of  Generalized Quartic Anharmonic Oscillators}

\author*[1]{W.H. Klink}\email{william-klink@uiowa.edu} 
\author*[2]{W. Schweiger}\email{wolfgang.schweiger@uni-graz.at} 

\affil*[1]{\orgdiv{Department of Physics and Astronomy}, \orgname{University of Iowa}, \orgaddress{\city{Iowa City}, \country{USA}}}

\affil*[2]{\orgdiv{Institute of Physics}, \orgname{University of Graz}, \orgaddress{\street{Universit\"atsplatz 5}, \city{Graz}, \postcode{A-8010}, \country{Austria}}}

\abstract{
This paper deals with the partial solution of the energy eigenvalue problem for generalized symmetric quartic oscillators. 
Algebraization of the problem is achieved by expressing the Schr\"odinger operator in terms of the generators of a nilpotent group, which we call the quartic group. Energy eigenvalues are then seen to depend on the values of the two Casimir operators of the group. This dependence exhibits a scaling law which follows from the scaling properties of the group generators.
Demanding that the potential gives rise to polynomial solutions in a particular Lie algebra element puts constraints on the four potential parameters, leaving only two of them free. For potentials satisfying such constraints at least one of the energy eigenvalues and the corresponding eigenfunctions can be obtained in closed analytic form {by pure algebraic means.} {With our approach we extend the class of quasi-exactly solvable quartic oscillators which have been obtained in the literature by means of the more common $\mathrm{sl}(2,\mathbb{R})$ algebraization.} {Finally we show, how solutions of the generalized quartic oscillator problem give rise to solutions for a charged particle moving in particular non-constant electromagnetic fields.}}


\maketitle

\vfill\break
%
%
%
\section{Introduction}\label{sec1}
{Quantum anharmonic oscillators play a prominent role if one wants to model physical phenomena in molecular, atomic, nuclear and  particle physics. Therefore a huge literature deals with solution methods for anharmonic oscillator problems. Whereas the harmonic oscillator is exactly solvable in the sense that all its energy eigenvalues can be obtained by pure algebraic means and the corresponding eigenfunctions are known in closed analytic form, this is not the case for anharmonic oscillators. Therefore either numerical methods and/or approximations have to be applied. An up-to-date exposition of approximate solution methods to quantum anharmonic oscillators is given in Ref.~\cite{Turbiner23}. But as it turns out, under some restrictions on the coefficients of (polynomial) anharmonic oscillators, at least a finite portion of the energy eigenvalues and the corresponding eigenfunctions can be found by algebraic means. The most prominent example is the one-dimensional Schr\"odinger operator with sextic anharmonic oscillator potential~\cite{Singh78,TurbinerU87,Turbiner88,Bender96}
\begin{equation}\label{eq:H6}
H_{(6)}=-\frac{d^2}{dx^2}+a x^6 + b x^4 + c x^2 + d\, , \qquad a>0\, .
\end{equation}
The ansatz
\begin{equation}\label{eq:ans6}
\psi_N^{(6)}(x)= x^k\, P_N(x^2)\,  e^{-\frac{\alpha}{4} x^4-\frac{\beta}{2} x^2}\, ,
\end{equation}
with $P_N(x^2)$, $N\in\mathbb{N}_0$, being an $N$-th order polynomial in $x^2$ and $k=0,1$, depending on whether one is interested in parity even or odd solutions, respectively, leads to an eigenvalue equation of the form
\begin{equation}\label{eq:evh6}
h_{(6)} \phi(y)=\epsilon\,\phi(y)
\end{equation}
for $\phi(y)=P_N(x^2)$. If the potential parameters $a$, $b$, $c$ and $d$ are taken as appropriate functions of the parameters $\alpha$, $\beta$, $k$ and $N$, which occur in the ansatz~(\ref{eq:ans6}), the second-order differential operator $h_{(6)}$ can be written as a second-degree polynomial in the $\mathrm{sl}(2,\mathbb{R})$ generators $J^+_N$, $J^0_N$ and $J^-_N$ which act on the space of polynomials with degree less or equal to $N$~\cite{Turbiner16}. As a consequence, the eigenvalue problem~(\ref{eq:evh6}) becomes a system of $(N+1)$ linear homogenous equations for the coefficients of $P_N(y=x^2)$ and thus $(N+1)$ eigenvalues and corresponding eigenfunctions can be found by pure algebraic means. Note that the sextic oscillators solvable in this way are (for fixed parity) a two-parameter family of functions depending on the two variables $\alpha$ and $\beta$ which show up in the exponential of the ansatz~(\ref{eq:ans6}). 

This kind of $\mathrm{sl}(2,\mathbb{R})$ algebraization, however, is not only restricted to the sextic anharmonic oscillator. It also works for any (one-dimensional) Schr\"odinger operator which, by an appropriate change of coordinates and a gauge rotation, can be transformed into an operator that is a second-degree polynomial in the $\mathrm{sl}(2,\mathbb{R})$ generators $J^+_N$, $J^0_N$ and $J^-_N$. A systematic and comprehensive investigation of  one-dimensional (radial) Schr\"odinger operators which admit an $\mathrm{sl}(2,\mathbb{R})$ algebraization can be found in Ref.~\cite{Turbiner16}. Interestingly, also the known exactly solvable potentials, like the harmonic oscillator or the Morse potential are amenable to this kind of approach. 

Unfortunately the usual quartic anharmonic oscillator, i.e. a=0 in Eq.~(\ref{eq:H6}), which is of much more physical interest than the sextic oscillator, resists an $\mathrm{sl}(2,\mathbb{R})$ algebraization. It, however, has been recognized that at least one eigenvalue and the corresponding eigenfunction can be obtained by algebraic means for the generalized (symmetric) quartic oscillator~\cite{Skala97,Znojil16,Quesne17,Quesne18}
\begin{equation}\label{eq:quartpot1}
 V_{(4)}^\mathrm{gen}(x)=V_0+A|x| +Bx^2 +C|x|^3 +D x^4\, , \qquad V_0, A, B, C,D\in\mathbb{R},\,D>0\, ,
\end{equation}
if the coefficients $V_0$,  $A$, $B$, $C$ and $D$ satisfy some restrictions.
Note that this potential, unlike the sextic oscillator in Eq.~(\ref{eq:H6}), is a non-analytic function at $x=0$. Since the potential is symmetric, it has again even and odd solutions. Therefore one can start to apply $\mathrm{sl}(2,\mathbb{R})$ algebraization, let us say, for $x>0$ in analogy to the sextic case~\cite{Quesne17,Quesne18}. Inserting an ansatz of the form
\begin{equation}\label{eq:ans4}
\psi_N^{(4)}(x)= \tilde{P}_N(x)\,  e^{-\frac{1}{3} |x|^3+\tilde\alpha x^2- \tilde\beta |x|}\, ,\qquad x > 0,
\end{equation}
with $\tilde{P}_N(x)$ an $N$-th order polynomial in $x$, into the energy-eigenvalue equation for the generalized quartic oscillator~(\ref{eq:quartpot1}) leads now directly to an eigenvalue equation for $\tilde{P}_N(x)$ with a \lq\lq reduced Hamiltonian\rq\rq\ $h_{(4)}$. Provided that the potential parameters $V_0$,  $A$, $B$, $C$ and $D$ are appropriate functions of $\tilde{\alpha}$, $\tilde{\beta}$ and $N$, $h_{(4)}$ can again be written as a second-degree polynomial in the $\mathrm{sl}(2,\mathbb{R})$ generators $J^+_N$, $J^0_N$ and $J^-_N$ which act on the space of polynomials with degree less or equal to $N$. As a consequence, the eigenvalue problem for $h_{(4)}$ becomes a system of $(N+1)$ linear homogenous equations for the coefficients of $\tilde{P}_N(x)$ and thus $(N+1)$ eigenvalues and corresponding eigenfunctions can be found by algebraic means. Parity even and odd eigenfunctions of the quartic oscillator eigenvalue problem on the whole real axis are then obtained by symmetric or antisymmetric continuation of the $x>0$ solutions to $x<0$. Thereby one has to satisfy Neumann (parity even) or Dirichlet (parity odd) boundary conditions at $x=0$, since the eigenfunctions and their first derivative should be continuous at $x=0$. These boundary conditions, in general, depend on the energy eigenvalue and relate the two parameters $\tilde\alpha$ and $\tilde\beta$. As a consequence one ends up with a class of generalized quartic oscillators for which one knows just one energy eigenvalue and the corresponding parity even or parity odd eigenfunction. This class depends on one free parameter (either $\tilde\alpha$ or $\tilde\beta$). Without referring to the \lq\lq hidden $\mathrm{sl}(2,\mathbb{R})$ symmetry\rq\rq , Sk\'ala  et al.~\cite{Skala97} and Znojil~\cite{Znojil16} got the same class of partially  solvable generalized quartic oscillators by inserting the ansatz~(\ref{eq:ans4}) into the Schr\"odinger equation, extracting a formal eigenvalue problem for the coefficients of $\tilde{P}_N(x)$ and relating the parameters $\tilde\alpha$ and $\tilde\beta$ by means of the continuity of the eigenfunctions and their first derivative at $x=0$. 

Following Refs.~\cite{Znojil16} and \cite{Quesne17,Quesne18} we will call any quantum mechanical problem quasi-exactly solvable, if a finite portion of the energy spectrum and its associated eigenfunctions can be found in closed analytic form by algebraic means. Some arguments for adopting this rather general meaning of quasi-exact solvability are given in the appendix of Ref.~\cite{Znojil16}. The nice monograph by Ushveridze~\cite{Ushveridze94} deals with a large class of quasi-exactly solvable models, but there non-analyticities like the one occurring in the generalized anharmonic oscillator~(\ref{eq:quartpot1}) (at $x=0$), are precluded. In Ref.~\cite{Turbiner16}, quasi-exact solvability is somewhat stricter and essentially characterized by $\mathrm{sl}(2,\mathbb{R})$ algebraization (in combination with certain analyticity properties of the potential). Some generalizations of this Lie-algebraic setting, leading to wave functions that can be expressed in terms of exceptional orthogonal polynomials, can be found in Refs.~\cite{Gomez06,Gomez07}.

In this paper we are looking for quasi-exactly solvable quartic oscillators by applying a different kind of algebraization which makes use of a nilpotent group that we will call the \lq\lq quartic group\rq\rq $\mathcal{Q}$. Like irreducible representations of the (nilpotent) Heisenberg group give rise to harmonic-oscillator type Schr\"odinger operators, irreducible representations of the quartic group are connected with generalized quartic oscillator problems~\cite{Klink94}. This already implies a number of general properties of generalized quartic oscillators, the most important being the structure and scaling properties of energy eigenvalues as functions of the Casimir invariants of the quartic group. We are mainly interested in the class of symmetric quartic oscillators of the form (\ref{eq:quartpot1}) which possess solutions that are the product of a polynomial times an exponential, similar to Eq.~(\ref{eq:ans4}). We call such solutions \lq\lq polynomial solutions\rq\rq\ for short. Instead of $x$ we, however, consider these solutions rather as functions of $X_2$, which is one of the generators of the quartic group. Our polynomial ansatz fixes the parameters $V_0$,  $A$, $B$, $C$ and $D$  in $V_{(4)}^\mathrm{gen}(x)$ in terms of 4 parameters and gives us simple recursion relations for the polynomial coefficients. Three of these four parameters label the irreducible representation of $\mathcal{Q}$, the remaining parameter fixes the strength of the linear potential term in the Schr\"odinger operator relativ to the higher order terms. The recursion relations imply that this (relative) strength parameter is determined by the highest power occurring in the polynomial ansatz, the three remaining parameters are again restricted by the requirement that the solution and its first derivative are continuous at $x=0$. In this way we end up with a two-parameter family of generalized quartic oscillators for which we know one energy eigenvalue and the corresponding parity even or parity odd eigenfunction. For these two free parameters one can take the two Casimir invariants of the quartic group. The requirement that a parity even and a parity odd solution arise from one and the same potential puts another constraint on the two open parameters and one ends up with a one-parameter family of generalized quartic oscillators for which one now knows two energy eigenvalues with the corresponding eigenfunctions having even and odd parity, respectively. 

In Sec.~\ref{sec2} we briefly review how irreducible representations of the Heisenberg group are related to the harmonic oscillator problem and it  is indicated how reducible representations give rise to the Hamiltonian of a charged particle in a constant magnetic field and to sublaplacian operators for the heat equation. The quartic group $\mathcal{Q}$ and its irreducible representations are then introduced in Sec.~\ref{sec3}. The generalized quartic oscillator Hamiltonian we are mainly interested in is specified in terms of generators of the quartic group and the algebra of these generators is exploited to derive scaling properties of the Hamiltonian and its eigenvalues and eigenfunctions. Polynomial solutions of the generalized quartic oscillator are discussed in Sec.~\ref{sec4}. General recursion relations for the polynomial coefficients and the constraints for parity even and odd solutions are derived. Explicit expressions for energy eigenvalues and corresponding eigenfunctions are given for polynomials of order 0, 1 and 2. Some results are also obtained for arbitrary order $N$ under the restriction that one of the Casimir invariants vanishes. Section~\ref{sec5} introduces (non-constant) electromagnetic fields associated with reducible representations of the quartic group. It is then demonstrated how solutions of the generalized quartic oscillator give rise to solutions for particles moving in such electromagnetic fields. Section~\ref{sec6} summarizes our results and outlines possible generalizations.}
%
%
%
\section{Review of the Harmonic Oscillator}\label{sec2}
The best known example of a relationship between a nilpotent group and an oscillator is that of the Heisenberg group and the harmonic oscillator~\cite{Klink94,Jorgensen85}.  The Heisenberg group is a nilpotent group that can be written as a matrix group with elements
\begin{eqnarray}\label{Heis gp}
(a, b_1, b_2)&:=&\left[\begin{array}{ccc}1&a&b_2\\0&1&b_1\\
0&0&1\\
\end{array}\right],
\end{eqnarray}
where $a, b_1, b_2 \in \mathbb{R}$.  Unitary irreducible representations are induced by $(0, b_1, b_2)\rightarrow e^{-i(\beta_1 b_1 +\beta_2 b_2)}$, where $\beta_1,\beta_2 \in \mathbb{R}$ are irreducible representation labels:
\begin{eqnarray}\label{Heis irrep}
(U_{a,b_1,b_2} ^{\beta_1,\beta_2} \phi)(x)=e^{-i(\beta_1 b_1+\beta_2(b_2 +b_1 x)} \phi(x+a) \, ,\qquad \phi\in L^2(\mathbb{R})\, .
\end{eqnarray}
Lie algebra representations are generated by one parameter subgroups:
\begin{subequations}\label{Heis alg}
\begin{align}
&(a,0,0)\rightarrow X_0 =i\frac{\partial}{\partial x}\, ,\\
&(0, b_1,0)\rightarrow X_1=\beta_1 + \beta_2 x\, , \\
&(0,0,b_2)\rightarrow X_2= \beta_2\, ,
\end{align}
\end{subequations}
with commutation relations $[X_0, X_1]=iX_2$, and all other commutators zero.

The harmonic oscillator Hamiltonian is a quadratic polynomial in Lie algebra elements:
\begin{eqnarray}\label{Heis HO}
H^{(\beta_1,\beta_2)} &=& X_0^2 + X_1^2\nonumber\\
&=& X_+ X_- +X_2,
\end{eqnarray}
where $X_{\pm} =X_0 \mp i X_1$\footnote{{Here and in the following we adopt a common convention and set $\hbar=2m=1$, with $m$ being the particle mass.}}.  The harmonic oscillator eigenfunctions can be obtained with raising and lowering operators, $[H^{(\beta_1,\beta_2)} , X_+]=X_2 X_+$, acting on the ground state $X_-\phi_0=0$.

Reducible representations of the Heisenberg group are obtained by inducing with the subgroup $(0,0,b_2)\rightarrow
e^{-i\beta_2 b_2} $, with Lie algebra elements given by
\begin{subequations}\label{red Heis}
\begin{align}
&X_0=i\frac{\partial}{\partial x}\, ,\\
&X_1= i\frac{\partial}{\partial y}+\beta_2 x\, , \\
&X_2= \beta_2\, ,
\end{align}
\end{subequations}
and Hamiltonian $H^{\beta_2}=X_0^2 +X_1^2= -\frac{\partial^2}{\partial x^2}+( i\frac{\partial}{\partial y}+\beta_2 x)^2$.
If a $z$ direction is added, this gives the Hamiltonian for a particle in a constant magnetic field with (dimensionless) strength
$\beta_2$ {(see Sec.~\ref{sec5})}.

If $H^{\beta_2}$ is Fourier transformed in $y$, the harmonic oscillator results, a property exploited by
Landau~\cite{Landau81} (Chap.~15), to get the eigenfunctions of a particle in an external {constant magnetic} field from eigenfunctions of the harmonic oscillator.  Group theoretically the Fourier transform decomposes the reducible representation of the Heisenberg group into a direct integral of irreducible representations. A more detailed discussion of this connection will be given in Sec.~\ref{sec5}.

Finally, the regular representation of the Heisenberg group (which is obtained by inducing with the identity element) can be used to define a sublaplacian and solve a heat equation.  The regular representation acts on elements of the Hilbert space
$L^2( G)$, $G=\mathbb{R}^3$ being the group manifold, as
\begin{eqnarray}\label{reg Heis}
(R_g F) ({{h}})&=& F({{h}} g)\, ,\qquad {{g, h \in G}}\nonumber,\\
(R_{(a,b_1,b_2)} F)(x,y,z)&=& F(x+a,y+b_1, z+b_2 +b_1 x) ,
\end{eqnarray}
with $F\in L^2 (G)$.  From this action Lie algebra elements are given by
\begin{subequations}\label{reg alg}
\begin{align}
&X_0=i\frac{\partial}{\partial x}\,  ,\\
&X_1=i\left(\frac{\partial}{\partial y}+x\frac{\partial}{\partial z}\right)\, ,\\
&X_2= i\frac{\partial}{\partial z},
\end{align}
\end{subequations}
and the Hamiltonian, now called a sublaplacian,  is
\begin{eqnarray}\label{subl}
\Delta&=&X_0^2 +X_1^2\nonumber\\
&=&-\frac{\partial^2}{\partial x^{\, 2}}-\left(\frac{\partial}{\partial y}+x\frac{\partial}{\partial z}\right)^2.
\end{eqnarray}
A great deal is known about sublaplacians of nilpotent groups (see Ref.~\cite{Jorgensen87}, Chap.~6) and in fact the (generalized) eigenfunctions of $\Delta$ are obtained by Fourier transforming in both $y$ and $z$ to get to the harmonic oscillator Hamiltonian.  The double Fourier transform decomposes the regular representation into a direct integral of irreducible representations, connected with the harmonic oscillator.  Using this fact makes it possible to solve the heat equation,
$\Delta p=\frac{\partial p}{\partial t}$, as first shown in Ref.~\cite{Hulanicki76}.  Using this structure it is also possible to solve the heat equation directly, as shown in Ref.~\cite{Jorgensen88}.

%
%
%
\section{The Quartic Group}\label{sec3}
Just as the Heisenberg group is intimately related to the harmonic oscillator, so too a group we call the quartic group, $\mathcal{Q}$, is intimately related to the (generalized) quartic anharmonic oscillator.  In this section we discuss the properties of the quartic group.  Its elements are written as
\begin{eqnarray}\label{quartic}
(a,\vec{b})=(a,b_1,b_2,b_3) :=\left[\begin{array}{cccc}1&a&\frac{a^2}{2}&b_3\\0&1&a&b_2\\
0&0&1&b_1\\0&0&0&1
\end{array}\right] \, , \qquad a,b_1,b_2,b_3\in \mathbb{R},
\end{eqnarray}
with the group operation given by
\begin{eqnarray}\label{gpop}
(a, \vec{b} ) (a^{\prime}, \vec{b}^{\,\prime})&=&(a+a^{\prime}, b_1 +b^{\prime}_1, b_2 +b^{\prime}_2+a b^{\prime}_1, b_3 +b^{\prime}_3 +ab^{\prime}_2 +\frac{a}{2}b^{\prime}_1)\, ,\\
(a, \vec{b})^{-1}&=&(-a, -b_1, -b_2 +ab_1, -b_3+ ab_2 - \frac{a^2}{2}b_1)\, .
\end{eqnarray}
The Heisenberg group is a subgroup of $\mathcal{Q}$ as can be seen by setting the parameter $b_1=0$.

The irreducible representations of $\mathcal{Q}$ can be obtained as induced representations, induced by the subgroup
$ (0, \vec{b})\rightarrow \pi ^{\vec{\beta} }(\vec{b} ):=e^{-i \vec{\beta}\cdot \vec{b} }$, ${\vec{\beta} \in \mathbb{R}^3}$. Then a unitary irreducible representation is given by
\begin{eqnarray}\label{Qirrep}
(U^{\vec{\beta}}_{(a,\vec{b})}\phi)(x)&=&e^{-i[\beta_1 b_1 +\beta_2(b_2 + b_1 x)+\beta_3(b_3 +b_2 x+ b_1 \frac{x^2}{2})]}
\phi(x+a)\, ,
\end{eqnarray}
with $(a,\vec{b}) \in \mathcal{Q}$, $\phi \in L^2 (\mathbb{R})$.

One parameter subgroups generate representations of the Lie algebra of~$\mathcal{Q}$:
\begin{subequations}\label{eq:generators}
\begin{align}
&(a,0\,0\,0)\rightarrow X_0 = i\frac{\partial}{\partial x},\\
&(0, b_1\, 0\, 0)\rightarrow X_1= \beta_1 +\beta_2 x +\beta_3 \frac{x^2}{2},\\
&(0,0\, b_2\, 0)\rightarrow X_2 = \beta_2 + \beta_3 x,\\
&(0,0\,0\, b_3)\rightarrow X_3  = \beta_3,
\end{align}
\end{subequations}
with commutation relations 
\begin{equation}
[X_0, X_1]=i X_2,\qquad [X_0, X_2]=iX_3,
\end{equation}
and all other commutators zero.

From these commutation relations it is seen that the Casimir operators are $X_3$ and
\begin{equation}\label{eq:casimir}
C:=2 X_1 X_3 -X_2^2\, .
\end{equation}
Hence the irreps can be labeled by $\beta_3$ and $c:=2\beta_1 \beta_3 -\beta_2^2$.  Representations with the same values of the Casimir operators are equivalent representations;  they are obtained by computing $\pi^{\vec{\beta}} ((a,\vec{0}) (0, \vec{b}) (-a,\vec{0}))=\pi^{\vec{\beta}_a }(0,\vec{b})$, so that representations $\vec{\beta} _a$ are equivalent to
$\vec{\beta}$ if
\begin{equation}\label{eq:trans}
\vec{\beta}_a = \left(\begin{array}{c}\beta_1 +a\beta_2 + \frac{1}{2}a^2 \beta_3\\
\beta_2 +a \beta_3\\ \beta_3\end{array}\right)\, .
\end{equation}

The automorphism group of the Lie algebra of $\mathcal{Q}$, {which preserves the commutation relations,} is given by
\begin{eqnarray}\label{aut gp}
\alpha_g (X_i) &=&g_{ij}  X_j \\
&&\hspace{-5.0cm}\hbox{with}\qquad g=\left[\begin{array}{cccc}g_{00}&g_{01}&g_{02}&g_{03}\\0&g_{11}&g_{12}&g_{13}\\
0&0&g_{00}g_{11}&g_{00}g_{12}\\0&0&0&g_{00}^2 g_{11}\\
\end{array}\right].
\end{eqnarray}

Associated with the automorphism group is the scaling operator, defined by
\begin{eqnarray}\label{scaling}
(S_t \phi)(x):=\sqrt{t}\, \phi( tx)\, , \quad t>0,\, \phi \in L^2 (\mathbb{R});
\end{eqnarray}
the factor $\sqrt{t}$ makes the scaling operator unitary.  The Lie algebra elements have definite scaling properties, namely
\begin{subequations}\label{scaling X}
\begin{align}
&S_t X_0 S_t^{-1}= t^{-1} X_0,\\
&S_t X_1 (\vec{\beta}) S_t^{-1} = t^{-1} X_1 (\vec{\beta}_t ),\\
&S_t X_2 (\vec{\beta} )  S_t^{-1} = t^{-2} X_2 (\vec{\beta}_t),\\
&S_t X_3 (\vec{\beta}) S_t ^{-1} = t^{-3} X_3 (\vec{\beta} _t),\\
{\hbox{with}\quad} & {\vec{\beta}_t :=( t\beta_1, t^2 \beta_2, t^3 \beta_3)\, .}
\end{align}
\end{subequations}

With this background we define the {generalized} quartic anharmonic oscillator Hamiltonian as
\begin{subequations}\label{eq:Hquart}
\begin{align}
H^{\vec{\beta}}_\alpha :=& \,X_0^2 + X_1^2 +\alpha X_2\\
=& -\frac{\partial}{\partial x}\frac{\partial}{\partial x}+(\beta_1 + \beta_2 x +\frac{\beta_3}{2} x^2 )^2 +\alpha (\beta_2 +\beta_3 x).
\end{align}
\end{subequations}
With appropriate values of $\alpha$ and $\beta_i$ this gives the usual quartic anharmonic oscillator Hamiltonian.  The eigenvalue problem to be solved is
\begin{eqnarray}\label{eq:eveq}
H^{\vec{\beta}}_\alpha \phi ^{\vec{\beta}}_n &=& E_n (\vec{\beta}) \phi ^{\vec{\beta}}_n .
\end{eqnarray}
From the definition of the Hamiltonian it follows that (see Eq.~(\ref{eq:trans}))
\begin{equation}
U_a H^{\vec{\beta}}_\alpha U_a^{-1} =H^{\vec{\beta}_a} _\alpha \, .
\end{equation}  
Applied to the eigenvalue problem this implies that 
\begin{equation}
U_a \phi^{\vec{\beta}} _n  =\phi ^{\vec{\beta}_a}_n
\end{equation} 
and the eigenvalues are functions of the Casimir invariants only.  Similarly 
\begin{equation}
S_t H^{\vec{\beta}}_\alpha S_t^{-1}=t^{-2} H^{\vec{\beta}_t} _\alpha \, ,\end{equation} 
from which it follows that 
\begin{equation}
S_t \phi^{\vec{\beta}}_n =\phi ^{\vec{\beta}_t}_n\end{equation} 
and the eigenvalue 
\begin{equation}
E_n (\vec{\beta} )=t^{-2} E_n (\vec{\beta}_t)\, .
\end{equation}  
Combining the invariance of the energy eigenvalues under both translation and scaling gives a functional equation of the form
 \begin{eqnarray}\label{energy1}
 E_n (t^3 \beta_3, t^4 c )&=&t^2 E_n (\beta_3, c),\\ \label{energy2}
 3\beta_3 \frac{\partial E_n}{\partial \beta_3}+ 4c \frac{\partial E_n}{\partial c}&=&2E_n;
 \end{eqnarray}%
 \noindent where Eq.~(\ref{energy2}) is obtained by differentiating both sides of Eq.~(\ref{energy1}) with respect to $t$ and then setting $t$ to one.  The solution of this first order partial differential equation is given by
\begin{eqnarray}\label{eq:escaling}
E_n&=& (\beta_3 )^{\frac{2}{3}} e_n(\frac{c^3}{\beta_3^4}),
\end{eqnarray}
so that the energy eigenvalues for a given irrep are given by a function, $e_n$, whose argument is a ratio of powers of Casimir invariants. {We want to emphasize that this result is not just restricted to the quasi-exact soluble models we will discuss in the following, but holds for any quartic oscillator Hamiltonian which has the structure (\ref{eq:Hquart}).}

%
 \section{Quasi-Exact Solutions of {Generalized}\\ Quartic Anharmonic Oscillator{s}}\label{sec4}
In this section we exhibit solutions for {generalized symmetric} quartic anharmonic {oscillators, Eq.~(\ref{eq:quartpot1}),} {which can be written as} $V = (X_1^2 +\alpha X_2)$ (see Eq.~(\ref{eq:Hquart})), where the  operators $X_1$ and $X_2$ are given in Eq.~(\ref{eq:generators}). 
{Comparing the potential {in Eq.~(\ref{eq:quartpot1})} with the potential $(X_1^2 +\alpha X_2)$ gives
 \begin{eqnarray}\label{eq:relZnojil}
 A&=&{2}\beta_1\beta_2+\alpha\beta_3\, ,\quad B=(\beta_1\beta_3+\beta_2^2)\, ,\quad C=\beta_2\beta_3\, , \nonumber\\
 {D}&=&\frac{\beta_3^2}{{4}}\quad \hbox{and}\quad {V_0}=\beta_1^2+\alpha\beta_2
\end{eqnarray}
for $x>0$. For $x<0$ one has to replace $\beta_1\rightarrow -\beta_1$ and $\beta_3\rightarrow -\beta_3$.}
The requirement of quasi-exact solvability leads then to restrictions for the choice of the potential parameters $\alpha$, $\beta_1$, $\beta_2$ and $\beta_3$.
That the solutions for the quartic oscillator are more complicated than {those for} the sextic oscillator arises from the asymptotic behavior of the two potentials.  Whereas {a normalizable solution of the} sextic oscillator goes asymptotically as $e^{-\gamma x^4}$ for $x\rightarrow \pm \infty$, {the asymptotic behavior of a normalizable} quartic oscillator  {solution is rather}  $e^{-\tilde{\gamma} x^3}$ for $x\rightarrow +\infty $ and $e^{\tilde{\gamma} x^3}$ for $x\rightarrow -\infty$. Here $\gamma>0$ and $\tilde{\gamma}>0$ are some appropriate constants depending on the strength of the sextic and quartic potential {terms}, respectively.  Thus, it will be necessary to find {normalizable} solutions separately for $x>0$ and $x<0$ {in case of the quartic oscillator} and then join them smoothly at $x=0$.

In Ref.~\cite{Znojil16} solutions for a potential of the form~(\ref{eq:quartpot1})
were found in the case of $V_0=0$ and $D=1$.  
Since $X_3=\beta_3$ is a Casimir invariant in our approach, we will not follow Ref.~\cite{Znojil16} {and set $D=1$}, but instead let $\beta_3$ have any positive {or negative real} value. Here it should be noted that the requirement of quasi-exact solvability {(and continuity of the solutions)} in Ref.~\cite{Znojil16} relates the potential strengths $A$, $B$ and $C$ so that they can finally be expressed in terms of only one free parameter. On the other hand, as we will see in the following, quasi-exact solvability will only restrict the parameters $\alpha$ and $\beta_2$ (or $\beta_1$) in our case, leaving still two parameters free.

 Our goal is to find solutions of the one-dimensional Schr\"odinger equation~(\ref{eq:eveq}) for Hamiltonians containing a potential of the form~(\ref{eq:quartpot1}). Phrased in our algebraic language, the Hamiltonian is written as
 \begin{equation}\label{eq:hamilton}
 H_\alpha^{\vec{\beta}}\! =\! \left\{\begin{array}{lcl}\hspace{-0.2cm}X_0^2 +X_1^2 +\alpha  X_2\! =
 \! -\frac{\partial^2}{\partial x^2}\! +\! (\beta_1 +\beta_2 x +\frac{\beta_3}{2}x^2 )^2\! +\alpha
 (\beta_2 +x\beta_3),&&\hspace{-0.1cm}x>0\, ,\\ &&\\
 \hspace{-0.2cm}\tilde{X}_0^2 +\tilde{X}_1^2 +\alpha  \tilde{X}_2 \! =
\!  -\frac{\partial^2}{\partial x^2}\! +\! (-\beta_1 +x\beta_2 -\frac{\beta_3}{2} x^2)^2\! +\alpha
 (\beta_2 -x\beta_3),&&\hspace{-0.1cm}x<0\, .\\
\end{array} \right. \end{equation}
In order to get a spatially symmetric potential  $V(x)=V(-x)$ one obviously has to employ different representations $X_i$ and $\tilde{X}_i$ of the quartic algebra. These representations differ just by the sign of $\beta_1$ and $\beta_3$. Changing the sign of $\beta_1$ and $\beta_3$
when going from $x>0$ to $x<0$ is obviously equivalent to taking $|x|$ for $x\in\mathbb{R}$ in potential terms containing odd powers of $x$ and leaving $\beta_1$ and $\beta_3$ untouched. Note that the representations of the quartic algebra used for $x>0$ and $x<0$ agree in the value $c$ of the Casimir $C$, but differ in the value $\beta_3$  of the Casimir $X_3$. They are therefore inequivalent. This, however, does not affect the scaling behavior~(\ref{eq:escaling}) of the energy eigenvalues, since scaling is determined only by even powers of $\beta_3$.

Since both, $X_i$ and $\tilde{X}_i$, satisfy the same algebra, the following considerations, which we make for $x>0$, will immediately apply to $x<0$ with
 $X_i$ replaced by $\tilde{X}_i$ or, equivalently, $\beta_1$ and $\beta_3$ by $-\beta_1$ and $-\beta_3$, respectively.
 Let us consider solutions of the form
\begin{equation}\label{eq:polsol}
\Psi_>(x)=p(x) e^{{\mp}\int dx\,X_1} \quad\hbox{with}\qquad \int dx\, X_1=\beta_1 x+\frac{\beta_2}{2}x^2+\frac{\beta_3}{6}x^3\quad\hbox{for}\quad \underline{x>0}\, .
\end{equation}
{Here and in what follows the upper (minus) sign in the exponential has to be taken for $\beta_3>0$, the lower (plus) sign for $\beta_3<0$.}
The integration constant in $\int dx\, X_1$ has been omitted, since it can be absorbed into the (unknown) function $p(x)$. The first two $x$-derivatives of $\Psi(x)$ are
 \begin{eqnarray}
 \Psi_>^{\prime} (x)&=&\left[p^{\prime} (x) {\mp} p(x) X_1 \right]\, e^{{\mp}\int dx\, X_1} \, ,\\
 \Psi_>^{\prime\prime} (x)&=&\left[p^{\prime\prime} (x) {\mp}2p^{\prime} (x) X_1 + p(x) (X_1^2{\mp}X_2)\right]\,e^{{\mp}\int dx\, X_1} \, .
 \end{eqnarray}
 The Schr\"odinger equation then to be solved for $x>0$ is
 \begin{eqnarray}\label{eq:eqp}
 -p^{\prime\prime}(x) {\pm}2 X_1\, p^{\prime}(x) +\left[({\pm}1+\alpha ) X_2- E\right]\,p(x)=0\, .
 \end{eqnarray}

As with the sextic oscillator we assume that $p$ is a polynomial.   However, unlike the sextic oscillator, we assume $p$ to be a polynomial in the Lie algebra element $X_2$,
 \begin{equation}
p(x)=\sum_{n=0}^N a_n\, X_2^n(x) \, ,
 \end{equation}
 so that $p^{\prime}(x)=\sum n\, a_n\, X_2^{n-1}\,X_3$ and $p^{\prime\prime}(x)= \sum n(n-1)\, a_n\,X_2^{n-2}\,X_3^2$. Substituting into Eq.~(\ref{eq:eqp}) gives
 \begin{eqnarray}\label{eq:recurs1}
&&\hspace{-1.8cm}-\sum_{n=2}^N a_n\,n(n-1)\,X_3^2\,X_2^{n-2}{\pm}\sum_{n=1}^N a_n\, n\, 2 X_1 X_3\,X_2^{n-1}\nonumber\\&&\hspace{3.0cm} + \,\, ({\pm}1+\alpha)\sum_{n=0}^N a_n\,X_2^{n+1}- E \sum_{n=0}^N a_n\,X_2^{n}=0\, .
\end{eqnarray}
 Now we express $2 X_1 X_3$ by means of the Casimir invariant $C=2X_1 X_3 -X_2^2$ (see Eq.~(\ref{eq:casimir})) so that
 the energy and coefficients in the polynomial become functions of Casimir invariants only.  Equation~(\ref{eq:recurs1}) thus becomes (after appropriate renaming of summation indices):
 \begin{eqnarray}
&&\hspace{-1.0cm}-\sum_{n=0}^{N-2} a_{n+2}\,(n+2)(n+1)\,X_3^2\,X_2^{n}{\pm}\sum_{n=0}^{N-1} a_{n+1}\, (n+1)\, C\,X_2^{n}\\
&&\hspace{0.0cm} {\pm}\,\,\sum_{n=2}^{N+1} a_{n-1}\,(n-1)\,X_2^{n}+ \,\, ({\pm}1+\alpha)\sum_{n=1}^{N+1} a_{n-1}\,X_2^{n}- E \sum_{n=0}^N a_n\,X_2^{n}=0\, .\nonumber
\end{eqnarray}
Demanding that the coefficient of $X_2^n$, $0\leq n\leq N+1$, should vanish, one ends up with a four-term recursion relation for the $a_n$s:
\begin{eqnarray}\label{eq:recurs2}
-(n+2)(n+1)\,\beta_3^2\,a_{n+2}{\pm}(n+1)\,c\, a_{n+1} -  E\, a_n +(\alpha {\pm}n)\, a_{n-1} =0\, .
\end{eqnarray}
Here we have used the abbreviation $c=2\beta_1 \beta_3 -\beta_2^2$ for the value of the Casimir $C$. The recursion relation
has to be understood such that $a_n=0$ if $n<0$ or $n>N$. For $n=N+1$ the recursion relation allows for a nonzero value of $a_N$ only if
$\alpha + N+1=0$;  for a given N this implies that
\begin{equation}\label{eq:alpha}
\alpha={\mp}(N+1)\, .
\end{equation}
Writing out the recursion relation for $n=0,1,2,\dots N$ gives $N+1$ linear equations for the coefficients $a_n$. Putting this system of equations in matrix form, one ends up with an eigenvalue problem
\begin{equation}\label{eq:eigenvec}
\mathcal{M}\, \vec{a} = E\, \vec{a}
\end{equation}
with the $(N+1)$-dimensional coefficient vector $\vec{a}=(a_0,a_1,a_2,\dots,a_N)^{\mathrm{T}}$ and the tridiagonal $(N+1)\times(N+1)$ Matrix $\mathcal{M}=(M_{n m})$ with matrix elements
\begin{eqnarray}\label{eq:matrixm}
M_{n (n-1)} &=&{\mp} (N+1-n)\, ,\nonumber\\ M_{n (n+1)} &=&{\pm} c (n+1)\, , \\ M_{n (n+2)} &=&- \beta_3^2 (n+2) (n+1)\, .\nonumber
\end{eqnarray}
All other matrix elements vanish. In order to obtain a non-trivial solution for $\vec{a}$, the eigenvalues $E$ are to be determined such that
\begin{equation}\label{eq:eigenv}
\det\left(\mathcal{M}-E\,\mathcal{I} \right)=0\, .
\end{equation}
These eigenvalues depend then, obviously, only on the values of the Casimirs $X_3$ and $C$.

The solutions for $\underline{x<0}$ are found in the same way. One just has to replace $X_i$ by $\tilde{X}_i$. With the ansatz
\begin{equation}
\Psi_<(x)=\tilde{p}(x) e^{{\mp}\int dx\,\tilde{X}_1} \, ,
\end{equation}
assuming $\tilde{p}(x)$ to be a polynomial in $\tilde{X}_2$,
\begin{equation}
\tilde{p}(x)=\sum_{n=0}^N a_n\, \tilde{X}_2^n(x) \, .
\end{equation}
The coefficients $a_n$ are seen to again satisfy the recursion relations~(\ref{eq:recurs2}), implying further that Eq.~(\ref{eq:eigenv}) leads to the same energy eigenvalues for $x>0$ and $x<0$. The reason is that the representations of the quartic algebra used for $x>0$ and $x<0$ are characterized by the same value of the Casimir $C$ and differ only in the sign of $\beta_3$. $\beta_3$, however, enters quadratically into the recursion relations.

The final step is now to match the solutions for $x>0$ and $x<0$ at $x=0$. To do this we notice first that the eigenfunctions must have definite parity properties, since our potential is spatially symmetric, $V(x)=V(-x)$. One can see immediately that
\begin{equation}\label{eq:parity}
\Psi(x)=\left\{\begin{array}{lll} \phantom{\pm}(\sum_{n=0}^N a_n\, X_2^n)\, e^{{\mp}\int dx X_1}&&x>0\\
& & \\
\pm(\sum_{n=0}^N a_n\, \tilde{X}_2^n)\, e^{{\mp}\int dx \tilde{X}_1}&&x<0\end{array}\right.
\end{equation}
is a parity even/odd (upper/lower sign) function which solves the Schr\"odinger equation for $x>0$ and $x<0$, if $E$ is a zero of the characteristic polynomial~(\ref{eq:eigenv}) and $\vec{a}$ is a solution of Eq.~(\ref{eq:eigenvec}). In order to be a solution of the Schr\"odinger equation on the whole real line, $\Psi(x)$ has to satisfy the continuity conditions
\begin{equation}
\lim_{\epsilon\rightarrow 0^+} \Psi(\epsilon)=\lim_{\epsilon\rightarrow 0^+}\Psi(-\epsilon)\quad\hbox{und}\quad \lim_{\epsilon\rightarrow 0^+} \Psi^\prime(\epsilon)=\lim_{\epsilon\rightarrow 0^+}\Psi^\prime(-\epsilon)\, .
\end{equation}
In the \underline{parity even case} $\Psi(x)$, as defined in Eq.~(\ref{eq:parity}), is already continuous at $x=0$. Continuity of the derivative at $x=0$ leads to the condition
\begin{equation}\label{eq:conteven}
a_0 \beta_1{\mp}\sum_{n=1}^N a_n\, (n\beta_3{\mp}\beta_1\beta_2)\, \beta_2^{n-1}=0\, .
\end{equation}
In the \underline{parity odd case} the derivative of $\Psi(x)$, as defined in Eq.~(\ref{eq:parity}), is already continuous at $x=0$. Continuity of $\Psi(x)$ at $x=0$ leads to the condition
\begin{equation}\label{eq:contodd}
\sum_{n=0}^N a_n\,\beta_2^n=0\, .
\end{equation}
These continuity conditions relate the three $\beta$s. As it turns out, apart from the $N=0$ even parity and $N=1$ odd parity cases, it is most convenient to fix $\beta_2$ and leave $\beta_1$ and $\beta_3$ as free parameters. Equivalently, one could also parameterize the potential by the values of the two Casimirs $\beta_3$ and $c$, respectively.

In principle this solves our problem. We are able to find at least one energy eigenvalue of the Hamilton operator~(\ref{eq:hamilton}) with the corresponding eigenfunction having the form~(\ref{eq:parity}). The formal procedure would be the following: First one has to solve the characteristic equation~(\ref{eq:eigenv}) to determine energy eigenvalues. These energy eigenvalue(s) are then inserted into Eq.~(\ref{eq:eigenvec}) to determine the coefficients $a_n$ (apart from one which provides the normalization of the wave function). Finally the continuity condition (\ref{eq:conteven}) or (\ref{eq:contodd}) is employed to fix $\beta_2$ such that the resulting parity even or odd solution solves the Sch\"odinger equation on the whole line.

In the following we will give examples {for $\underline{\beta_3>0}$}, starting with the simplest case $N=0$. We then proceed to $N=1,2$ and even try to find solutions for general $N$. {At the end of this section we will also comment on the $\beta_3<0$ case.}\\

\begin{figure}[t!]
\begin{centering}
\includegraphics[width=0.45\textwidth]{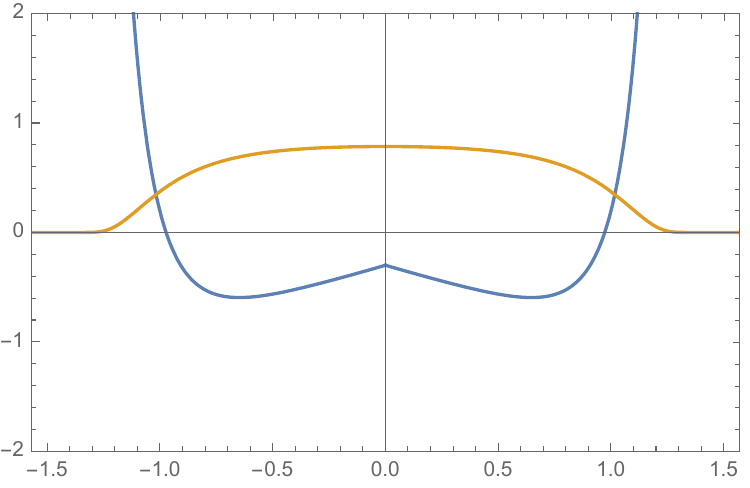}\hfill
\includegraphics[width=0.45\textwidth]{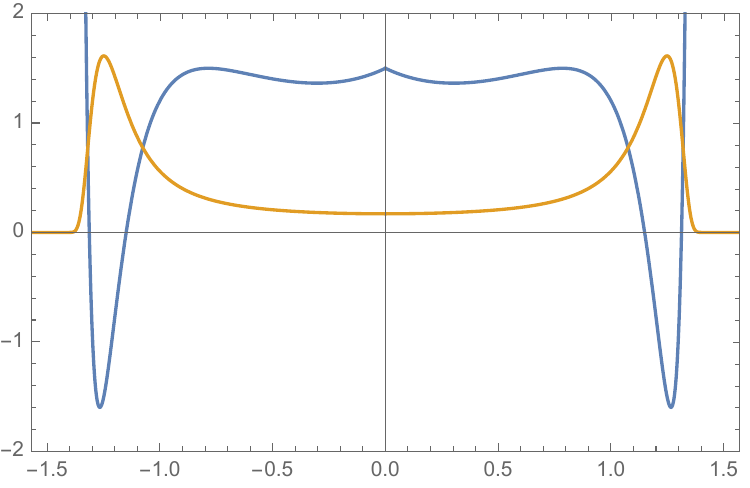}
\par\end{centering}
\caption{The potential $(X_1^2+\alpha X_2)$ for $\alpha=-1$, $\beta_1=0$, $\beta_2=0.3$, $\beta_3=0.6$ (left figure) and $\alpha=-1$, $\beta_1=0$, $\beta_2=-1.5$, $\beta_3=1$ (right figure) along with the corresponding (even parity) ground-state wave functions $\Psi_0$. Potential and wave functions are plotted as functions of $y=\arctan x$. The normalization $a_0$ of the wave function has been chosen such that $\int_{-\pi/2}^{\pi/2} dy\,\Psi_0^2(x(y))=1$.  \label{fig:N0even}} 
\end{figure}
\noindent\underline{$N=0$} ($\alpha=-1$):\\
In the $N=0$ case the \underline{parity-even} function (\ref{eq:parity}) can be written in the compact form
\begin{equation}
\Psi_0(x) = a_0\, e^{-\left(\beta_1 |x| +\frac{\beta_2}{2} x^2+\frac{\beta_3}{6}|x|^3\right)}\, .
\end{equation}
This function is an \underline{$E=0$} solution.
The continuity condition (\ref{eq:conteven}) implies then that \underline{$\beta_1=0$}. There is no non-trivial parity odd solution in this  case, since the continuity condition~(\ref{eq:contodd}) would immediately imply that $a_0=0$. In the parity even case we are thus left with $\beta_2$ and $\beta_3$ as free parameters. $\beta_3$ {is} positive, $\beta_2$ can be either positive or negative. For $\beta_3>0$ the slope of the potential in the limit $x\rightarrow 0^{\pm}$ is $\mp \beta_3/2$. Therefore it is at least a double well potential, but can even be more complicated as Fig.~\ref{fig:N0even} shows. The corresponding $E=0$ eigenfunctions are also plotted in Fig.~\ref{fig:N0even}. Since they do not exhibit a node, they are ground state wave functions. For the energy eigenvalue $E=0$ the scaling behavior (\ref{eq:escaling}), which describes the dependence of $E$ on the parameters $\beta_i$ is trivially satisfied.\\

\begin{figure}[t!]
\begin{centering}
\includegraphics[width=0.45\textwidth]{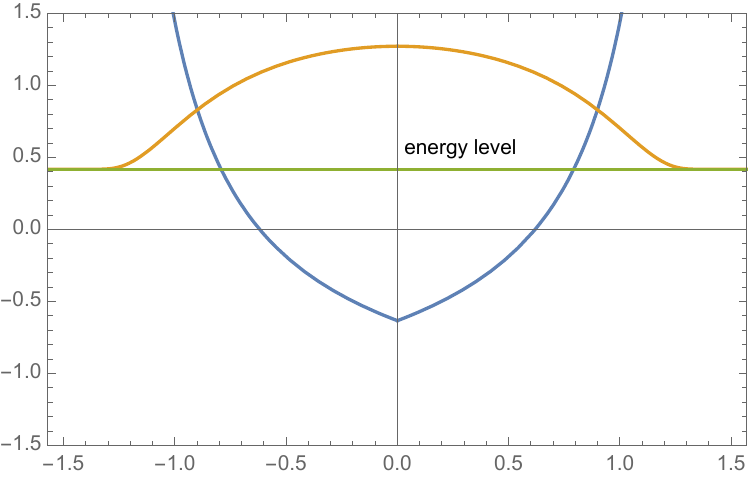}\hfill
\includegraphics[width=0.45\textwidth]{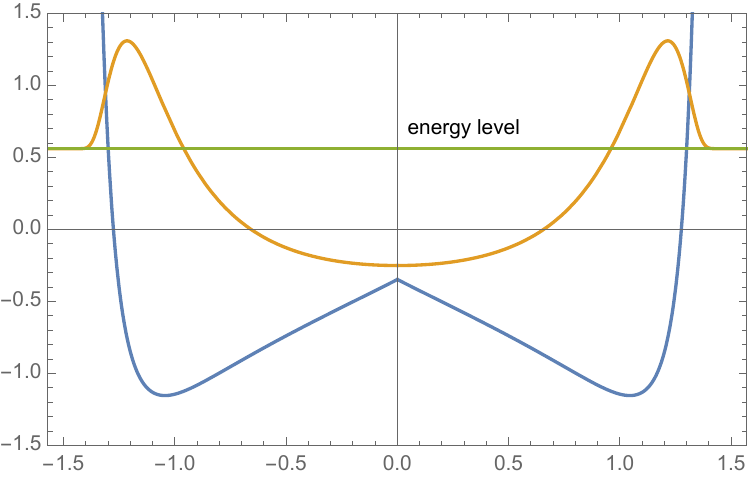}
\par\end{centering}
\caption{The potential $(X_1^2+\alpha X_2)$ for $\alpha=-2$, $\beta_1=0.7$, $\beta_3=0.1$ (left figure) and $\alpha=-2$, $\beta_1=-0.7$, $\beta_3=0.1$ (right figure) along with the corresponding (even parity) wave functions $\Psi_1^+$. $\beta_2$ is fixed acoording to Eq.~(\ref{eq:N1b2}). Potential and wave functions are plotted as functions of $y=\arctan x$. The normalization $a_1$ of the wave function has been chosen such that $\int_{-\pi/2}^{\pi/2} dy\,{\Psi_1^+}^2(x(y))=1$.  \label{fig:N1even}}
\end{figure}

\noindent\underline{N=1} ($\alpha=-2$):\\
In the $N=1$ case the \underline{parity-even} function
\begin{equation}
\Psi_1^+(x) = \left(a_0+a_1 (\beta_2+\beta_3 |x|)\right)\, \Psi_0(x)\, ,
\end{equation}
with $a_0=- E\,a_1$ solves the Schr\"odinger equation for
\begin{equation}\label{eq:N1b2}
E={\beta_1^2-\frac{\beta_3}{2\beta_1} }\qquad\hbox{if}\qquad \beta_2=\beta_1^2+\frac{\beta_3}{2\beta_1}\, .
\end{equation}
Potentials and corresponding wave functions $\Psi_1^+$ for two parameter sets are plotted in Fig.~\ref{fig:N1even}.
The derivative of the potential in the limit  $x\rightarrow 0^{\pm}$ is $\mp {2} (\beta_3-\beta_1 \beta_2)$. With $\beta_2$ given by Eq.~(\ref{eq:N1b2}) it becomes $\pm {(2\beta_1^3-\beta_3)}$. This means that one obtains a potential of the anharmonic oscillator type for ${(2\beta_1^3-\beta_3)}>0$ (minimum at $x=0$) and a double well for ${(2\beta_1^3-\beta_3)}<0$ (local maximum at $x=0$). Interestingly, the corresponding wave function $\Psi_1^+$ is a ground-state wave function (no node) for the anharmonic oscillator, whereas it is the wave function of a second excited state (two nodes) for the double well.\footnote{We have checked our results numerically by means of {\tt Mathematica} using the build-in function {\tt NDEigensystem} with Dirichlet boundary conditions. To do this we have transformed the real line $-\infty<x<\infty$ to the finite interval $-\frac{\pi}{2}\leq y\leq\frac{\pi}{2}$ by setting $x=\tan y$. The numerical values for the energies agree with our analytical results up to 6 digits. For the double well, e.g., {\tt Mathematica} gives $E_0=-0.732365$, $E_1=-0.366215$ and $E_2=0.561429$. $E_2$ is in perfect agreement with our analytical result.}

Expressing the energy eigenvalue~(\ref{eq:N1b2}) in terms of the Casimirs, as in Eq.~(\ref{eq:escaling}), to exhibit its scaling behavior, it takes on the form
\begin{equation}\label{eq:scalingN1}
E=\beta_3^{\frac{2}{3}}\, e\left(\frac{c^3}{\beta_3^4}\right) \qquad\hbox{with}\qquad e(\xi)=\pm\,(-\xi)^{\frac{1}{6}}\,\, .
\end{equation}
where \lq\lq$+$\rq\rq\ has to be taken for $E>0$ and \lq\lq$-$\rq\rq\ for $E<0$.

\begin{figure}[t!]
\begin{centering}
\includegraphics[width=0.45\textwidth]{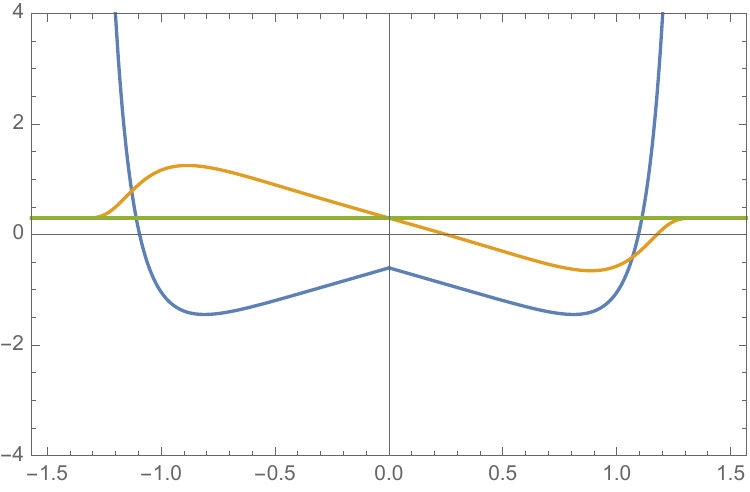}\hfill
\includegraphics[width=0.45\textwidth]{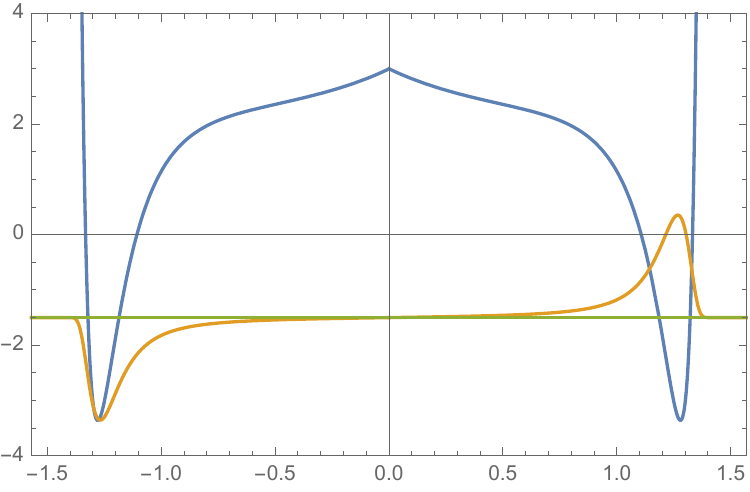}
\par\end{centering}
\caption{The potential $(X_1^2+\alpha X_2)$ for $\alpha=-2$, $\beta_1=0$, $\beta_2=0.3$, $\beta_3=0.6$ (left figure) and $\alpha=-2$, $\beta_1=0$, $\beta_2=-1.5$, $\beta_3=1$ (right figure) along with the corresponding (odd parity) wave functions $\Psi_1^-$. Potential and wave functions are plotted as functions of $y=\arctan x$. The normalization $a_1$ of the wave function has been chosen such that $\int_{-\pi/2}^{\pi/2} dy\,{\Psi_1^-}^2(x(y))=1$.  \label{fig:N1odd}}
\end{figure}

The $N=1$ \underline{parity-odd} solution, corresponding to the energy eigenvalue
\begin{equation}
E={\beta_2}\ ,
\end{equation}
is
\begin{equation}
\Psi_1^-(x) = \mathrm{sign}(x)\,\Psi_1^+(x)\, ,
\end{equation}
with $\mathrm{sign}(x)$ denoting the sign function. Continuity at $x=0$ implies that $\beta_1=0$ so that $\beta_2$ and $\beta_3>0$ are left as free parameters. In Fig.~\ref{fig:N1odd} we have plotted the potential together with the $E={\beta_2}$ eigenfunctions for the two sets of parameters which we have already used in the $N=0$ even parity case. {The} slope of the potential for $x\rightarrow 0^\pm$ is $\mp {2}\beta_3$ and thus the potential has a local maximum at $x=0$. Both eigenfunctions exhibit one node which means that they represent the lowest lying odd parity state and hence the first excited state of the spectrum.

\noindent It is easily checked that $E={\beta_2}$ exhibits the same scaling behavior (cf. Eq.~(\ref{eq:scalingN1})) as the $N=1$ parity even solution.\\

\noindent\underline{N=2} ($\alpha=-3$):\\
Also for $N=2$ the energy eigenvalue equation (\ref{eq:eigenv}) together with the continuity conditions (\ref{eq:conteven}) and (\ref{eq:contodd}) for parity even and parity odd solutions, respectively, can be solved analytically. In the \underline{parity even} case one finds two pairs $(E,\beta_2)$ of real solutions for the energy eigenvalue equation (\ref{eq:eigenv}) and the continuity condition (\ref{eq:conteven}), namely
\begin{eqnarray}\label{eq:EN2even}
E&=&{2}\frac{\beta_1^3-3\beta_3\mp\sqrt{\beta_1^6-6\beta_1^3\beta_3+4\beta_3^2}}{5\beta_1}\, ,\nonumber\\
\beta_2&=&\frac{7 \beta_1^3+4\beta_3\pm ^3\!\!\!\sqrt{\beta_1^6-6\beta_1^3\beta_3+4\beta_3^2}}{10\beta_1}\, .
\end{eqnarray}
The condition $(\beta_1^6-6\beta_1^3\beta_3+4\beta_3^2)\geq 0$ guarantees that $E$ and $\beta_2$ are real. The (even parity) eigenfunction corresponding to the energy eigenvalue $E$ is
\begin{equation}
\Psi_2^+(x) = \left(a_0+a_1 (\beta_2+\beta_3 |x|)+a_2 (\beta_2+\beta_3 |x|)^2\right)\, \Psi_0(x)
\end{equation}
with $a_0=( E^2{/2} - \beta_2^2 + 2 \beta_1 \beta_3) a_2$ and $a_1=- E a_2$. Note that one has to take either the upper or the lower sign for the roots in Eq.~(\ref{eq:EN2even}).

On the other hand, there is just one real solution of the energy eigenvalue equation (\ref{eq:eigenv}) and the continuity condition~(\ref{eq:contodd}) in the \underline{parity odd} case, namely
\begin{equation}\label{eq:EN2odd}
E={4} \beta_1^2 \qquad\hbox{and}\qquad \beta_2=\frac{4 \beta_1^3 + \beta_3}{2 \beta_1}
\end{equation}
giving rise to the eigenfunction $\Psi_2^-(x)=\mathrm{sign}(x)\Psi_2^+(x)$.

\begin{figure}[t!]
\begin{centering}
\includegraphics[width=0.45\textwidth]{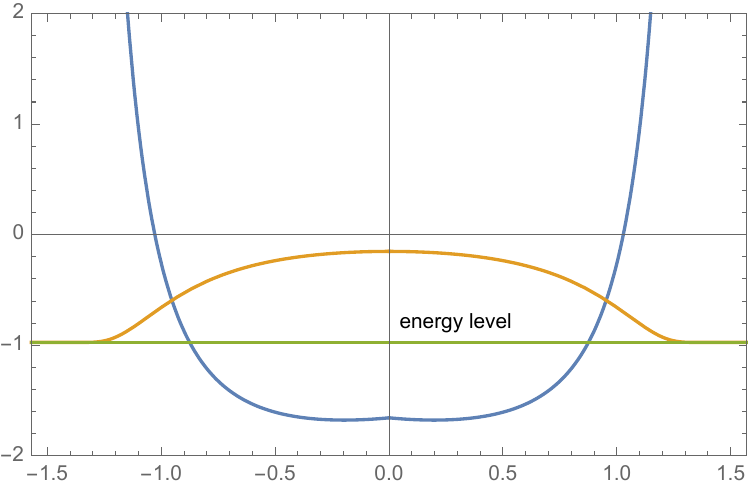}\hfill
\includegraphics[width=0.45\textwidth]{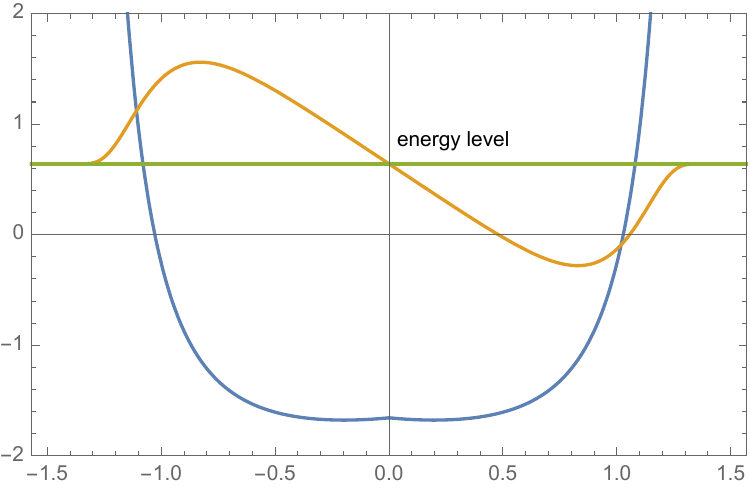}
\par\end{centering}
\caption{The potentials $(X_1^2+\alpha X_2)$ for $\alpha=-3$, $\beta_1=0.4$, $\beta_3=\frac{4}{7} (2 +3 \sqrt{2})\beta_1^3$ and $\beta_2$ chosen according to Eqs.~(\ref{eq:EN2even}) and (\ref{eq:EN2odd})  along with the corresponding even and odd parity wave functions $\Psi_2^+$ (left figure) and $\Psi_2^-$ (right figure), respectively. Potential and wave functions are plotted as functions of $y=\arctan x$. The normalization $a_2$ of the wave function has been chosen such that $\int_{-\pi/2}^{\pi/2} dy\,{\Psi_2^\pm}^2(x(y))=1$. \label{fig:N2}}
\end{figure}

The interesting point is now that $\beta_2$ in Eq.~(\ref{eq:EN2even}) and Eq.~(\ref{eq:EN2odd}) can be made equal, if
\begin{equation}
\beta_3=\frac{4}{7} (2 \pm 3 \sqrt{2}) \beta_1^3\, .
\end{equation}
In order that $\beta_3>0$, one has to take the upper sign if $\beta_1>0$, otherwise the lower sign. This means that this particular choice of $\beta_3$ gives rise to a potential (which still contains $\beta_1$ as free parameter), for which we know two energy eigenvalues with corresponding parity even and parity odd eigenfunctions, respectively. This situation is plotted in Fig.~\ref{fig:N2}, where the  energies given in Eqs. (\ref{eq:EN2even}) and (\ref{eq:EN2odd}) for $\beta_1=0.4$ represent the ground state and first excited state with corresponding even and odd parity eigenfunctions. For $\beta_1=-0.4$ one would get the second and third excited state of the corresponding potential.

For $N=2$ the energy eigenvalue equation~(\ref{eq:eigenv}) reads
\begin{equation}
E^3+{4} c E+{4}\beta_3^2=0\, .
\end{equation}
Writing the solution of this cubic equation  in the form~(\ref{eq:escaling}), one can read off the scaling behavior of the energy eigenvalues:
\begin{eqnarray}
E&=&\beta_3^{2/3}\, e\left(\frac{c^3}{\beta_3^4}\right)\qquad\hbox{with}\nonumber\\ e(\xi)&=&\frac{2^{{5/3}} \xi^{1/3}}{3 \left(1-\sqrt{1+\frac{16}{27}\xi}\right)^{1/3}} - 2^{{1/3} }\left(1-\sqrt{1+\frac{16}{27}\xi}\right)^{1/3} \, .
\end{eqnarray}
Note that this holds for the parity even, Eq.~(\ref{eq:EN2even}), as well as for the parity odd solution, Eq.~(\ref{eq:EN2odd}), since the energy eigenvalue equation~(\ref{eq:eigenv}) just depends on the values $c$ and $\beta_3$ of the Casimirs.\\

\noindent\underline{$N>2$}:\\
From what we have seen, it becomes more and more complicated with increasing $N$ to find analytic solutions $(E,\beta_2)$ of the energy-eigenvalue equation (\ref{eq:eigenv}) and the continuity condition (\ref{eq:conteven}) or (\ref{eq:contodd}) for the corresponding eigenfunctions. Surprisingly, it is possible (by means of {\tt Mathematica}) to find solutions for $N=3$, but the expressions for $E$ and $\beta_2$ in terms of $\beta_1$ and $\beta_3$ tend to become rather lenghty, in particular for the parity even case. Thus one may consider putting some restrictions on the potential parameters $\beta_i$ so that the energy-eigenvalue equation and the continuity conditions become simpler. Looking  at the recursion relation~(\ref{eq:recurs2}), an obvious simplification is achieved if we demand that the value of the Casimir $C$ vanishes, i.e. \underline{$c=0$}. In this case the four-term recursion relation is reduced to a three-term recursion relation and the $(N+1)\times(N+1)$-matrix $\mathcal{M}$ in Eq.~(\ref{eq:eigenvec}) becomes a bidiagonal matrix. One can see now that
\begin{equation}\label{eq:Mc0}
\mathcal{M}\,\vec{a}=\vec{0}\, ,
\end{equation}
with $\mathcal{M}$ given by Eq.~(\ref{eq:matrixm}) ($c=0$), has a non-trivial solution $\vec{a}\neq\vec{0}$, if $N\neq 2+3k$, $k\in\mathbb{N}_0$. This means that \underline{$E=0$} is an eigenvalue for the allowed values of $N$, provided that the corresponding eigenfunctions satisfy either of the continuity conditions (\ref{eq:conteven}) or (\ref{eq:contodd}), respectively.
The wave-function coefficients $a_n$ are most easily calculated by means of the downward recursion relation
\begin{equation}\label{eq:down}
a_{n-3}=-\frac{n (n-1)}{N-n+3}\,  {\beta_3^2}\, a_n\, ,
\end{equation}
starting with $n=N$. {Equation~(\ref{eq:down}) follows immediately from Eq.~(\ref{eq:recurs2}) by taking $c=E=0$, $\alpha=-(N+1)$.} It is also understood that $a_n=0$ for $n<0$. {The condition $c=0$ implies that $\beta_1=\frac{\beta_2^2}{2\beta_3}$.} The parity-even continuity condition (\ref{eq:conteven}) or the parity-odd continuity condition  (\ref{eq:contodd}) restrict finally the possible values of the potential parameter $\beta_2$, leaving only {$\beta_3\neq 0$} as  free parameter. {One can easily check that solutions of Eq.~(\ref{eq:Mc0}) satisfy either of the continuity conditions (\ref{eq:conteven}) or (\ref{eq:contodd}) automatically, if $\beta_1=\beta_2=0$ and $N=3 k$ or $N=3k+1$, $k\in\mathbb{N}_0$, respectively. }
{A further} check with {\tt Mathematica} shows that at least one of the continuity conditions has a real solution for $\beta_2\neq 0$, if $N=1,3,4,6,7$. For {(the allowed)} $N>7$ the continuity conditions are only satisfied by setting $\beta_1=\beta_2=0$. The possible choices of $\beta_2$, up to $N=10$, which lead to an $E=0$ eigenvalue in the case of vanishing Casimir $c=0$ are summarized in Tab.~\ref{tab:ezero}.
\begin{table}[t!]
\begin{tabular}{|c|c|c|} \hline
$N$&parity even & parity odd \\
\hline
0 &  0 & $\times$\\
1 &  $2^{1/3} \beta_3^{2/3}$ & $0$\\
2 & $\times$ & $\times$\\
3 &  $0\, ,$ $2 \beta_3^{2/3}$ & $2^{1/3} \beta_3^{2/3}$\\
4 &  $2^{1/3} (3\pm\sqrt{7})^{1/3} \beta_3^{2/3}$ & $0\, ,$ $2 \beta_3^{2/3}$\\
5 & $\times$ & $\times$\\
6 &  $0\, ,$ $(11\pm\sqrt{51})^{1/3}\beta_3^{2/3}$ & $(5\pm\sqrt{15})^{1/3}\beta_3^{2/3}$\\
7 &  $\times$ & $0\, ,$ $(7\pm\sqrt{21})^{1/3}\beta_3^{2/3}$\\
8 & $\times$ & $\times$\\
9 & 0 & $\times$\\
10 & $\times$ & 0\\ 
\hline
\end{tabular}
\caption{Possible choices of the potential parameter $\beta_2$ which give rise to a {generalized} quartic potential with $E=0$ energy eigenvalue in the case of $c=0$ ($\beta_1=\frac{\beta_2^2}{2\beta_3}$).}\label{tab:ezero}
\end{table}
\begin{figure}[t!]
\begin{centering}
\includegraphics[width=0.45\textwidth]{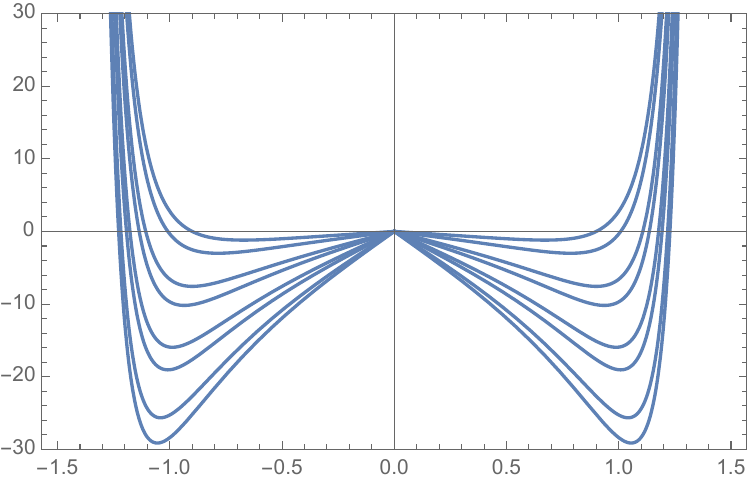}\hfill
\includegraphics[width=0.55\textwidth]{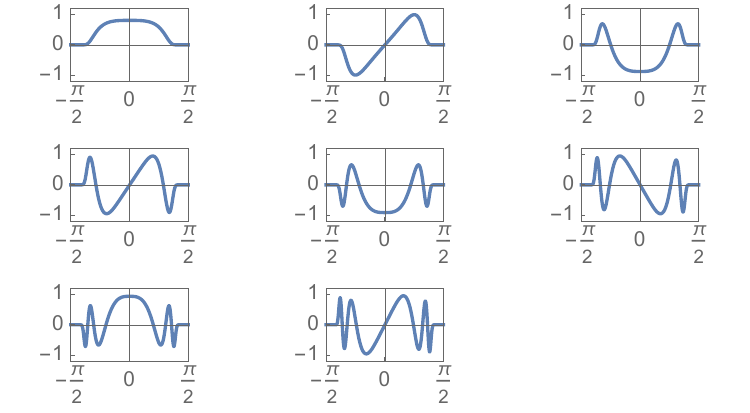}
\par\end{centering}
\caption{{The potential~(\ref{eq:skala}) for $N=0,1,3,4,6,7,9,10$ (left) along with the corresponding $E=0$ wave functions (right). Potentials become deeper and the number wave-function nodes increases with increasing $N$.} Potential and wave functions are plotted as functions of $y=\arctan x$. The normalization  of the wave functions has been chosen such that $\int_{-\pi/2}^{\pi/2} dy\,{\Psi_N^+}^2(x(y))=1$. \label{fig:Ngen}}
\end{figure}

{Let us now consider the case $\beta_1=\beta_2=0$, $\beta_3=2$, $\alpha=-(N+1)$ which has been discussed in some detail in Ref.~\cite{Skala97}. This choice of parameters leads to the potentials 
\begin{equation}\label{eq:skala}
V_N(x)=-2(N+1) |x|+x^4\, , \qquad N=3k, 3k+1\, , \quad k=0,1,2,3,..
\end{equation}
These are double-well potentials of increasing depth which are plotted in Fig.~\ref{fig:Ngen} along with the corresponding wave functions for $N=0,1,3,4,6,7,9,10$. What happens is that for $N=0$ the $E=0$ eigenfunction corresponds to the (parity even) ground state. For $N=1$ the depth increases, the ground state goes down and the first (parity odd) excited state (with 1 node) now lies at $E=0$. With increasing $N$ the potential becomes deeper and deeper and the number of wave function nodes of the zero energy solution increases. By making the potential deeper, the energy levels go down and at the allowed values of $N$ one of the (excited) levels just crosses $E=0$. This also leads to the observed alternating pattern of parity even and parity odd $E=0$ eigenfunctions. Our findings for the potential~(\ref{eq:skala}) and corresponding $E=0$ wave functions agree with those in Ref.~\cite{Skala97}.
}

{Let us finally consider the \underline{$\beta_3<0$} case. A closer inspection of the $\beta_3<0$ case now reveals that the polynomial ansatz~(\ref{eq:parity}) (with the lower sign) just provides the same class of quasi-integrable potentials as the $\beta_3>0$ case. There is a one-to-one correspondence between the $\beta_3>0$ and the $\beta_3<0$ cases which just consists in reversing the sign of all the potential parameters. The replacement $(\beta_1,\beta_2,\beta_3,\alpha)\rightarrow (-\beta_1,-\beta_2,-\beta_3,-\alpha)$ does not change the potential (see, e.g., Eqs.~(\ref{eq:quartpot1}) and (\ref{eq:relZnojil})) and hence neither the energy spectrum nor the shape of the energy eigenfunctions. In the functional form of the energy eigenfunctions~(\ref{eq:parity}) the sign change of the potential parameters is accompanied by a sign change of the coefficients $a_n$, $n$ odd, which can be traced back to  Eq.~(\ref{eq:matrixm}).}

%
%
%
 \section{The Electromagnetic Field Related to the\\ Quartic Anharmonic Oscillator}\label{sec5}
 Associated with every oscillator given by some nilpotent group is an electromagnetic field {problem} (see Ref.~\cite{Jorgensen87}, Chap.~7;  also Ref.~\cite{Jorgensen85}).  {By inducing with} the subgroup $(0,0,b_2,b_3)\rightarrow e^{-i(\beta_2 b_2 +\beta_3 b_3)} $ {one ends up with a reducible representation of the quartic group.}  The resulting generators are now given by
  \begin{subequations}
  \begin{align}
  &(a,0\,0\,0)\rightarrow X_0 = i\frac{\partial}{\partial x},\\
&(0, b_1\, 0\, 0)\rightarrow X_1= i\frac{\partial}{\partial y} +\beta_2 x +\beta_3 \frac{x^2}{2},\\
&(0,0\, b_2\, 0)\rightarrow X_2 = \beta_2 + \beta_3 x,\\
&(0,0\,0\, b_3)\rightarrow X_3  = \beta_3,
 \end{align}
 \end{subequations}
 with Hamiltonian
 \begin{eqnarray}
 H^{(\beta_2,\beta_3)}_\alpha&=&X_0^2+ X_1^2 +\alpha X_2\nonumber\\
 &=& -\frac{\partial^2}{\partial x^2} +( i \frac{\partial}{\partial y}+\beta_2 x +\beta_3 \frac{x^2}{2})^2+ \alpha (\beta_2 + x\beta_3) {.}
 \end{eqnarray}
{By adding a kinetic energy term $P_z^2$ for a particle which moves freely in $z$ direction one ends up with the Hamiltonian
\begin{eqnarray}
H_{\mathrm{em}}:={H}_\alpha^{(\beta_2,\beta_3)}\otimes {I}_z\, \oplus\, {I}_{x\,y}\otimes {P}_z^2 && \\ && \hspace{-3.2cm}=
-\frac{\partial^2}{\partial x^2}+(-i\frac{\partial}{\partial y}-\beta_2 x-\beta_3 \frac{x^2}{2})^2-\frac{\partial^2}{\partial z^2}+\alpha (\beta_2+\beta_3 x)  \nonumber\, ,
\end{eqnarray}
where $I_{xy}$ and $I_z$ are unity operators acting on the $(x,y)$ and $z$ coordinates, respectively.
This Hamiltonian describes a particle in an electromagnetic field 
\begin{equation}\label{eq:emfield}
\vec{E}(\vec{r})=\left(\begin{array}{c} -\alpha\beta_3\\0\\0 \end{array}\right)\, , \qquad
\vec{B}(\vec{r})=\left(\begin{array}{c} 0\\0\\\beta_2+\beta_3 x \end{array}\right)\, ,
\end{equation}
the corresponding electrodynamical potential being
\begin{equation}
\left(A^\mu(\vec{r})\right)=\left(\alpha(\beta_2+\beta_3 x),0,\beta_2 x+\beta_3\frac{x^2}{2},0\right)\, .
\end{equation}
This means that, starting with the quartic group instead of the Heisenberg group allows us to study not just the case of a charged particle in a constant magnetic field, but gives rise to a more general electromagnetic field configuration.
The energy-eigenvalue problem for $H_{\mathrm{em}}$,
\begin{eqnarray}\label{eq:EVem2}
\left[-\frac{\partial^2}{\partial x^2}+(-i\frac{\partial}{\partial y}-\beta_2 x-\beta_3 \frac{x^2}{2})^2-\frac{\partial^2}{\partial z^2}+\alpha (\beta_2+\beta_3 x) \right] \Phi_E(x,y,z)&&\\ &&\hspace{-2.2cm} = E\,\Phi_E(x,y,z)\, ,\nonumber
\end{eqnarray}
can now be related to the (one-dimensional) quartic oscillator problem by switching to a mixed configuration- momentum-space representation, which is obtained by means of a Fourier transformation in the $y$ and $z$ variables:
\begin{eqnarray}\label{eq:phimixed}
\tilde{\Phi}_{E}(x,p_y,p_z)&=&\frac{1}{2\pi}\,\int dy\,dz\,e^{-i p_y y-ip_z z}\, \Phi_{E}(x,y,z)\, .
\end{eqnarray}
This mixed wave function satisfies a differential equation in the $x$ variable which has the form
\begin{eqnarray}
\left[-\frac{\partial^2}{\partial x^2}+(p_y-\beta_2 x-\beta_3 \frac{x^2}{2})^2+p_z^2+\alpha (\beta_2+\beta_3 x) \right] \tilde{\Phi}_E(x,p_y,p_z)&&\\ &&\hspace{-1.7cm} = E\,\tilde{\Phi}_E(x,p_y,p_z)\, .\nonumber
\end{eqnarray}
With $\lambda=E-p_z^2$ and $\beta_1=-p_y$ this is just the eigenvalue problem~(\ref{eq:eveq}) for the generalized quartic oscillator. This means that, knowing a solution $\Phi_E(x,y,z)$ of the electromagnetic field eigenvalue problem~(\ref{eq:EVem2}), its Fourier transformation $\tilde{\Phi}_E(x,p_y,p_z)$ in the $y$ and $z$ variable (see Eq.~(\ref{eq:phimixed})) gives rise to a solution $\Psi_\lambda(x)$ of the anharmonic oscillator problem by setting
\begin{equation}
\Psi_{\lambda}(x)=\tilde{\Phi}_E(x,p_y,p_z)\,\,\hbox{with}\,\, \beta_1=-p_y\, ,\,\,\lambda=E-p_z^2\,\,\hbox{and}\,\,p_y\, , p_z \,\,\hbox{fixed}\, .
\end{equation} 
Note that each $p_y$ is associated with a different anharmonic oscillator problem. The electromagnetic field problem corresponds to a reducible representation of the quartic algebra, the quartic oscillator problem rather to an irreducible one. 
The Hamiltonian of the electromagnetic problem is a direct integral of Hamiltonians ${H}^{(\beta_1,\beta_2,\beta_3)}_\alpha$ for the (one-dimensional) quartic oscillator problem, i.e.
\begin{eqnarray}
{H}_{\mathrm{em}}&=&{H}_\alpha^{(\beta_2,\beta_3)}\otimes {I}_z\, \oplus\, {I}_{x\,y}\otimes {P}_z^2\nonumber\\
&=&\left(\int_{\mathbb{R}}^\oplus dp_y\,{H}_\alpha^{(\beta_1=-p_y,\beta_2,\beta_3)}\right)\otimes {I}_z\, \oplus\, {I}_{x\,y}\otimes {P}_z^2\, .
\end{eqnarray}
Correspondingly the eigenfunctions of the electromagnetic field problem can be decomposed into eigenfunctions of the anharmonic oscillator problem
\begin{eqnarray}\label{eq:efdecomp}
\Phi_E(x,y,z)&=&\frac{1}{2\pi} \int dp_y\,dp_z\,  e^{i p_y y+i p_z z} \tilde{\Phi}_E(x,p_y,p_z)\, ,\nonumber\\
&=& \frac{1}{2\pi} \int dp_y\,dp_z\,  e^{i p_y y+i p_z z} {\Psi}_{\lambda}(x)\, \delta(\lambda-E+p_z^2)\, ,\nonumber\\
\end{eqnarray}
with $\beta_1=-p_y$.
Note that $\lambda=\lambda(\alpha,\beta_1=-p_y,\beta_2,\beta_3)$ is a function of the integration variable $p_y$. Equation~(\ref{eq:efdecomp}) shows, how eigenfunctions and eigenvalues of the one-dimensional anharmonic oscillator problem and a corresponding three-dimensional electromagnetic field problem are related, provided that the boundary conditions in $x$ direction are the same. Since $[{H}_q,{P}_y]=[{H}_q,{P}_z]=0$, one can look for simultaneous eigenfunctions of $H_{\mathrm{em}}$, ${P}_y$ and ${P}_z$. These are then obviously of the form (see Eq.~(\ref{eq:efdecomp}))
\begin{equation}\label{eq:pwave}
\Phi_{E p_y p_z}(x,y,z)=\frac{1}{2\pi} e^{i p_y y+i p_z z} \Psi_\lambda(x)\,\, \hbox{with}\,\, \beta_1=-p_y\, \hbox{and}\,\lambda=E-p_z^2\, .
\end{equation}
Here the plane waves have been normalized to a pure delta function.

Up to this point  our considerations hold for general electromagnetic fields of the form~(\ref{eq:emfield}).Using the quartic oscillator solutions derived in Sec.~\ref{sec4} for constructing special solutions of the electromagnetic field problem by means of Eq.~(\ref{eq:efdecomp}) or Eq.~(\ref{eq:pwave}), one must take into account that quasi-integrability puts constraints on either $\beta_1$ or $\beta_2$. For the $N=0$ parity even and the $N=1$ parity odd cases one has $\beta_1=0$, which means that only  $p_y=\beta_1=0$ solutions of the electromagnetic field problem can be constructed from the known quartic oscillator solutions. For the other cases  $\beta_2$ becomes a function of $\beta_1$ and $\beta_3$ which has to be taken care of in Eq.~(\ref{eq:efdecomp}) or Eq.~(\ref{eq:pwave}).
}

%
%


\section{{Summary} and Outlook}\label{sec6}
{It is known from previous work~\cite{Skala97,Znojil16,Quesne17,Quesne18} that the energy-eigenvalue problem for the generalized symmetric quartic anharmonic oscillator, given in Eq.~(\ref{eq:quartpot1}), is quasi exactly solvable. One energy eigenvalue and the corresponding parity even or parity odd eigenfunction can be calculated by algebraic means. The deeper reason is that this quartic oscillator problem admits an $\mathrm{sl}(2,\mathbb{R})$ algebraization like the sextic oscillator~\cite{Turbiner16}, as shown in Refs.~\cite{Quesne17,Quesne18}. 

In the present work we have attempted another kind of algebraization by means of a nilpotent group, the quartic group $Q$. We have shown that certain generalized quartic oscillator problems can be associated with irreducible representations of the quartic group by expressing the corresponding Hamiltonian in terms of generators of the quartic group (see Eq.~(\ref{eq:Hquart})). In this way the potential parameters in Eq.~(\ref{eq:quartpot1}) become functions of the three irreducible representation labels $\beta_1\, ,\beta_2\, ,\beta_3$ and a further parameter $\alpha$ which essentially fixes the relative strength of the linear potential term as compared to the higher order terms. For general quartic oscillator Hamiltonians of this kind, which include e.g. also the usual quartic oscillator, we were able to derive the structure and scaling properties of energy eigenvalues as functions of the Casimir invariants of the quartic group. 

In the sequel we have looked for generalized symmetric quartic oscillators~(\ref{eq:quartpot1}) which give rise to parity even and parity odd polynomial solutions of the form~(\ref{eq:polsol}) with the polynomial being a function of the generator $X_2$ rather than $x$. For the  quartic oscillator~(\ref{eq:quartpot1}) the solutions for $x>0$ and $x<0$ have to be treated separately and the corresponding Hamiltonians belong to irreducible representations which differ in the sign of the parameters $\beta_1$ and $\beta_3$. With a polynomial ansatz of order $N$ the Schr\"odinger equation reduces to an $(N+1)$-dimensional algebraic eigenvalue equation for the polynomial coefficients, provided that the relative strength parameter takes the integer value $\alpha=\mp (N+1)$ (depending on whether $\beta_3\gtrless 0$). One of the $\beta$-parameters becomes a function of the other two $\beta$s, if the $x>0$ and $x<0$ solutions are smoothly matched at $x=0$ such that one ends up with a parity-even or parity-odd energy eigenfunction. Since the continuity condition at $x=0$ depends on the energy eigenvalue, we have finally obtained a class of generalized symmetric quartic oscillators which is parameterized by the discrete parameter $\alpha$ and two continuous parameters for which one can take the values of the two Casimir invariants of the quartic group. For this class of anharmonic oscillators one knows one energy eigenvalue with the corresponding parity even or parity odd eigenfunction. For $N=|\alpha-1|\geq 2$ it was even possible to satisfy the continuity condition for parity even and odd solutions at the same time by appropriately fixing two of the three $\beta$s in terms of the remaining one. In this way we got a one-parameter family of symmetric quartic oscillators for which one now knows two energy eigenvalues, one belonging to a parity even and one to a parity odd eigenfunction, respectively. Explicit expressions for energy eigenvalues and corresponding eigenfunctions in terms of the free $\beta$ parameter(s) have been derived for $N=0\, , 1\, , 2$. These cases were discussed in some detail and the scaling law of the energy eigenvalues in terms of the Casimir invariants, which was derived in  Sec.~\ref{sec3}, has been verified. Under the simplifying assumption that the eigenvalue $c=2 \beta_1 \beta_3-\beta^2_2$ of one of the Casimirs becomes zero, we were able to find sets of potential parameters for $N>2$ which give rise to an $E=0$ solution of the energy eigenvalue problem. The one-parameter family of potentials determined by the parameter set $\beta_1=\beta_2=0$, $\beta_3\gtrless 0$, e.g., provides an $E=0$  eigenvalue, if $N\neq2+k$, $k\in \mathbb{N}_0$. The coefficients of the corresponding eigenfunctions are given by a simple two-term recursion relation. Furthermore we saw that generalized quartic oscillator potentials  with $c=2 \beta_1 \beta_3-\beta^2_2=0$, $\beta_2, \beta_1 \neq 0$ and $E=0$ energy eigenvalue exist only for $N=0\,,1\,,3\,, 4\,, 6\,, 7$. The class of quasi-exactly solvable generalized quartic oscillator potentials which we found by means of our approach covers those already known from Refs.~\cite{Skala97,Znojil16,Quesne17,Quesne18}, but includes also new ones, e.g. those for which one can calculate two eigenvalues with corresponding even and odd parity eigenfunction. Also the potentials which we found for $c=0$,  
$\beta_2, \beta_1 \neq 0$ and $N>2$ which provide an $E=0$ energy eigenvalue are, to the best of our knowledge, new. 
Finally we have shown, how reducible representations of the quartic group give rise to Hamiltonians describing the movement of a charged particle in certain non-constant electromagnetic fields and how solutions of the quartic oscillator can  be  used to find solutions of the corresponding electromagnetic field problem.

All of these ideas can be generalized to higher power polynomial potentials, such as the (generalized) sextic anharmonic oscillator.  For the sextic oscillator there is a corresponding sextic nilpotent group, whose irreducible representations can be used to write the sextic anharmonic oscillator Hamiltonian in terms of sextic Lie algebra elements.  Quasi-exactly solvable sextic oscillators obtained by means of $\mathrm{sl}(2,\mathbb{R})$ algebraization provide either positive or negative-parity algebraic solutions.  It will be interesting to see, whether our kind of approach leads also to quasi-exactly solvable (generalized) sextic oscillators for which part of the parity even as well as parity odd eigenfunctions can be calculated by algebraic means. This will be the focus of future investigations.}

\section*{Acknowledgements}
The authors acknowledge the financial support by the University of Graz.

\vfill\break


\begin{thebibliography}{19}
\ifx \bisbn   \undefined \def \bisbn  #1{ISBN #1}\fi
\ifx \binits  \undefined \def \binits#1{#1}\fi
\ifx \bauthor  \undefined \def \bauthor#1{#1}\fi
\ifx \batitle  \undefined \def \batitle#1{#1}\fi
\ifx \bjtitle  \undefined \def \bjtitle#1{#1}\fi
\ifx \bvolume  \undefined \def \bvolume#1{\textbf{#1}}\fi
\ifx \byear  \undefined \def \byear#1{#1}\fi
\ifx \bissue  \undefined \def \bissue#1{#1}\fi
\ifx \bfpage  \undefined \def \bfpage#1{#1}\fi
\ifx \blpage  \undefined \def \blpage #1{#1}\fi
\ifx \burl  \undefined \def \burl#1{\textsf{#1}}\fi
\ifx \doiurl  \undefined \def \doiurl#1{\url{https://doi.org/#1}}\fi
\ifx \betal  \undefined \def \betal{\textit{et al.}}\fi
\ifx \binstitute  \undefined \def \binstitute#1{#1}\fi
\ifx \binstitutionaled  \undefined \def \binstitutionaled#1{#1}\fi
\ifx \bctitle  \undefined \def \bctitle#1{#1}\fi
\ifx \beditor  \undefined \def \beditor#1{#1}\fi
\ifx \bpublisher  \undefined \def \bpublisher#1{#1}\fi
\ifx \bbtitle  \undefined \def \bbtitle#1{#1}\fi
\ifx \bedition  \undefined \def \bedition#1{#1}\fi
\ifx \bseriesno  \undefined \def \bseriesno#1{#1}\fi
\ifx \blocation  \undefined \def \blocation#1{#1}\fi
\ifx \bsertitle  \undefined \def \bsertitle#1{#1}\fi
\ifx \bsnm \undefined \def \bsnm#1{#1}\fi
\ifx \bsuffix \undefined \def \bsuffix#1{#1}\fi
\ifx \bparticle \undefined \def \bparticle#1{#1}\fi
\ifx \barticle \undefined \def \barticle#1{#1}\fi
\bibcommenthead
\ifx \bconfdate \undefined \def \bconfdate #1{#1}\fi
\ifx \botherref \undefined \def \botherref #1{#1}\fi
\ifx \url \undefined \def \url#1{\textsf{#1}}\fi
\ifx \bchapter \undefined \def \bchapter#1{#1}\fi
\ifx \bbook \undefined \def \bbook#1{#1}\fi
\ifx \bcomment \undefined \def \bcomment#1{#1}\fi
\ifx \oauthor \undefined \def \oauthor#1{#1}\fi
\ifx \citeauthoryear \undefined \def \citeauthoryear#1{#1}\fi
\ifx \endbibitem  \undefined \def \endbibitem {}\fi
\ifx \bconflocation  \undefined \def \bconflocation#1{#1}\fi
\ifx \arxivurl  \undefined \def \arxivurl#1{\textsf{#1}}\fi
\csname PreBibitemsHook\endcsname

\bibitem[\protect\citeauthoryear{Turbiner and del
  Valle~Rosales}{2023}]{Turbiner23}
\begin{bbook}
\bauthor{\bsnm{Turbiner}, \binits{A.V.}},
\bauthor{\bsnm{Valle~Rosales}, \binits{J.C.}}:
\bbtitle{Quantum Anharmonic Oscillator}.
\bpublisher{World Scientific}, \blocation{Singapore}
(\byear{2023})
\doiurl{10.1142/13252} 
\end{bbook}
\endbibitem

\bibitem[\protect\citeauthoryear{Singh et~al.}{1978}]{Singh78}
\begin{barticle}
\bauthor{\bsnm{Singh}, \binits{V.}},
\bauthor{\bsnm{Biswas}, \binits{S.N.}},
\bauthor{\bsnm{Datta}, \binits{K.}}:
\batitle{{The anharmonic oscillator and the analytic theory of continued
  fractions}}.
\bjtitle{Phys. Rev. D}
\bvolume{18},
\bfpage{1901}
(\byear{1978})
\doiurl{10.1103/PhysRevD.18.1901}
\end{barticle}
\endbibitem

\bibitem[\protect\citeauthoryear{Turbiner and Ushveridze}{1987}]{TurbinerU87}
\begin{barticle}
\bauthor{\bsnm{Turbiner}, \binits{A.V.}},
\bauthor{\bsnm{Ushveridze}, \binits{A.G.}}:
\batitle{Spectral singularities and quasi-exactly solvable quantal problem}.
\bjtitle{Physics Letters A}
\bvolume{126}(\bissue{3}),
\bfpage{181}--\blpage{183}
(\byear{1987})
\doiurl{10.1016/0375-9601(87)90456-7}
\end{barticle}
\endbibitem

\bibitem[\protect\citeauthoryear{Turbiner}{1988}]{Turbiner88}
\begin{barticle}
\bauthor{\bsnm{Turbiner}, \binits{A.V.}}:
\batitle{{Quasiexactly solvable problems and SL(2) group}}.
\bjtitle{Commun. Math. Phys.}
\bvolume{118},
\bfpage{467}
(\byear{1988})
\doiurl{10.1007/BF01466727}
\end{barticle}
\endbibitem

\bibitem[\protect\citeauthoryear{Bender and Dunne}{1996}]{Bender96}
\begin{barticle}
\bauthor{\bsnm{Bender}, \binits{C.M.}},
\bauthor{\bsnm{Dunne}, \binits{G.V.}}:
\batitle{{Quasiexactly solvable systems and orthogonal polynomials}}.
\bjtitle{J. Math. Phys.}
\bvolume{37},
\bfpage{6}--\blpage{11}
(\byear{1996})
\doiurl{10.1063/1.531373}
\end{barticle}
\endbibitem

\bibitem[\protect\citeauthoryear{Turbiner}{2016}]{Turbiner16}
\begin{barticle}
\bauthor{\bsnm{Turbiner}, \binits{A.V.}}:
\batitle{{One-dimensional quasi-exactly solvable Schr\"odinger equations}}.
\bjtitle{Phys. Rept.}
\bvolume{642},
\bfpage{1}--\blpage{71}
(\byear{2016})
\doiurl{10.1016/j.physrep.2016.06.002}
\end{barticle}
\endbibitem

\bibitem[\protect\citeauthoryear{Sk{\'a}la et~al.}{1997}]{Skala97}
\begin{barticle}
\bauthor{\bsnm{Sk{\'a}la}, \binits{L.}},
\bauthor{\bsnm{Dvo{\v r}{\'a}k}, \binits{J.}},
\bauthor{\bsnm{Kapsa}, \binits{V.}}:
\batitle{{Analytic solutions of the Schr\"odinger equation for the modified
  quartic oscillator}}.
\bjtitle{Int. J. Theor. Phys.}
\bvolume{36}(\bissue{12}),
\bfpage{2953}--\blpage{2961}
(\byear{1997})
\doiurl{10.1007/BF02435720}
\end{barticle}
\endbibitem

\bibitem[\protect\citeauthoryear{Znojil}{2016}]{Znojil16}
\begin{barticle}
\bauthor{\bsnm{Znojil}, \binits{M.}}:
\batitle{Symmetrized quartic polynomial oscillators and their partial exact
  solvability}.
\bjtitle{Physics Letters A}
\bvolume{380}(\bissue{16}),
\bfpage{1414}--\blpage{1418}
(\byear{2016})
\doiurl{10.1016/j.physleta.2016.02.035}
\end{barticle}
\endbibitem

\bibitem[\protect\citeauthoryear{Quesne}{2017}]{Quesne17}
\begin{barticle}
\bauthor{\bsnm{Quesne}, \binits{C.}}:
\batitle{{Quasi-exactly solvable symmetrized quartic and sextic polynomial
  oscillators}}.
\bjtitle{The European Physical Journal Plus}
\bvolume{132}(\bissue{11}),
\bfpage{450}
(\byear{2017})
\doiurl{10.1140/epjp/i2017-11718-y}
\end{barticle}
\endbibitem

\bibitem[\protect\citeauthoryear{Quesne}{2018}]{Quesne18}
\begin{barticle}
\bauthor{\bsnm{Quesne}, \binits{C.}}:
\batitle{{Quasi-exactly solvable polynomial extensions of the quantum harmonic
  oscillator}}.
\bjtitle{J. Phys. Conf. Ser.}
\bvolume{1071}(\bissue{1}),
\bfpage{012016}
(\byear{2018})
\doiurl{10.1088/1742-6596/1071/1/012016}
\end{barticle}
\endbibitem

\bibitem[\protect\citeauthoryear{Ushveridze}{1994}]{Ushveridze94}
\begin{bbook}
\bauthor{\bsnm{Ushveridze}, \binits{A.G.}}:
\bbtitle{Quasi-Exactly Solvable Models in Quantum Mechanics}.
\bpublisher{Taylor \& Francis}, \blocation{New York}
(\byear{1994})
\burl{https://books.google.at/books?id=u4jv1bydQXMC}
\end{bbook}
\endbibitem

\bibitem[\protect\citeauthoryear{Gomez-Ullate et~al.}{2007a}]{Gomez06}
\begin{barticle}
\bauthor{\bsnm{Gomez-Ullate}, \binits{D.}},
\bauthor{\bsnm{Kamran}, \binits{N.}},
\bauthor{\bsnm{Milson}, \binits{R.}}:
\batitle{Quasi-exact solvability in a general polynomial setting}.
\bjtitle{Inverse Problems}
\bvolume{23},
\bfpage{1915}--\blpage{1942}
(\byear{2007})
\doiurl{10.1088/0266-5611/23/5/008}
\end{barticle}
\endbibitem

\bibitem[\protect\citeauthoryear{Gomez-Ullate et~al.}{2007b}]{Gomez07}
\begin{barticle}
\bauthor{\bsnm{Gomez-Ullate}, \binits{D.}},
\bauthor{\bsnm{Kamran}, \binits{N.}},
\bauthor{\bsnm{Milson}, \binits{R.}}:
\batitle{Quasi-exact solvability beyond the sl(2) algebraization}.
\bjtitle{Physics of Atomic Nuclei}
\bvolume{70}(\bissue{3}),
\bfpage{520}--\blpage{528}
(\byear{2007})
\doiurl{10.1134/S1063778807030118}
\end{barticle}
\endbibitem

\bibitem[\protect\citeauthoryear{Klink}{1994}]{Klink94}
\begin{bchapter}
\bauthor{\bsnm{Klink}, \binits{W.H.}}:
\bctitle{{Nilpotent groups and anharmonic oscillators}}.
In: \beditor{\bsnm{Tanner}, \binits{E.}},
\beditor{\bsnm{Wilson}, \binits{R.}} (eds.)
\bbtitle{Noncompact Groups and Some of Their Applications}.
\bsertitle{NATO ASI series C},
vol. \bseriesno{429},
pp. \bfpage{301}--\blpage{313}
(\byear{1994})
\end{bchapter}
\endbibitem

\bibitem[\protect\citeauthoryear{J{\o}rgensen and Klink}{1985}]{Jorgensen85}
\begin{barticle}
\bauthor{\bsnm{J{\o}rgensen}, \binits{P.E.T.}},
\bauthor{\bsnm{Klink}, \binits{W.}}:
\batitle{{Quantum mechanics and nilpotent groups. I. The curved magnetic
  field}}.
\bjtitle{Publ. Res. Inst. Math. Sci.}
\bvolume{21},
\bfpage{969}--\blpage{999}
(\byear{1985})
\end{barticle}
\endbibitem

\bibitem[\protect\citeauthoryear{Landau and Lifshitz}{1981}]{Landau81}
\begin{bbook}
\bauthor{\bsnm{Landau}, \binits{L.D.}},
\bauthor{\bsnm{Lifshitz}, \binits{E.M.}}:
\bbtitle{{Quantum Mechanics: Non-Relativistic Theory}}.
\bsertitle{Course of Theoretical Physics},
vol. \bseriesno{3}.
\bpublisher{Butterworth-Heinemann}
(\byear{1981})
\end{bbook}
\endbibitem

\bibitem[\protect\citeauthoryear{J{\o}rgensen}{1987}]{Jorgensen87}
\begin{bbook}
\bauthor{\bsnm{J{\o}rgensen}, \binits{P.E.T.}}:
\bbtitle{{Operators and Representation Theory.}}
\bpublisher{North Holland}, \blocation{Amsterdam}
(\byear{1987}).
\end{bbook}
\endbibitem

\bibitem[\protect\citeauthoryear{Hulanicki}{1976}]{Hulanicki76}
\begin{barticle}
\bauthor{\bsnm{Hulanicki}, \binits{A.}}:
\batitle{{The distribution of energy in the Brownian motion in the Gaussian
  field and analytic-hypoellipticity of certain subelliptic operators on the
  Heisenberg group}}.
\bjtitle{Studia Mathematica}
\bvolume{56},
\bfpage{165}--\blpage{173}
(\byear{1976})
\doiurl{10.4064/sm-56-2-165-173}
\end{barticle}
\endbibitem

\bibitem[\protect\citeauthoryear{J{\o}rgensen and Klink}{1988}]{Jorgensen88}
\begin{barticle}
\bauthor{\bsnm{J{\o}rgensen}, \binits{P.E.T.}},
\bauthor{\bsnm{Klink}, \binits{W.H.}}:
\batitle{{Spectral transform for the sub-Laplacian on the Heisenberg group}}.
\bjtitle{Journal d'Analyse Math{\'e}matique}
\bvolume{50}(\bissue{1}),
\bfpage{101}--\blpage{121}
(\byear{1988})
\doiurl{10.1007/BF02796116}
\end{barticle}
\endbibitem

\end{thebibliography}

\end{document}